\pdfoutput=1
\RequirePackage{ifpdf}
\ifpdf 
\documentclass[pdftex]{sigma}
\else
\documentclass{sigma}
\fi
\usepackage{booktabs} 
\usepackage{amsmath}  
\usepackage{array}    

\usepackage{mathtools}
\numberwithin{equation}{section}

\newtheorem*{Theorem*}{Theorem}

\theoremstyle{definition}

\begin{document}

\renewcommand{\PaperNumber}{***}

\FirstPageHeading

\ShortArticleName{Applications to Economic Growth}

\ArticleName{Simultaneous Holotheticity and Bi-Hamiltonian Structures in Economic Growth Theory}

\Author{Sarah FINKLE~$^{\rm a}$ and Roman G. SMIRNOV~$^{\rm b}$\footnote{The corresponding author.}}

\AuthorNameForHeading{S.~Finkle and R.G.~Smirnov}

\Address{$^{\rm a), b)}$~Department of Mathematics and Statistics,  Dalhousie University, 6297 Castine Way, PO BOX 15000, Halifax, Nova Scotia, B3H 4R2, Canada} 

\Email{\mail{SarahFinkle@dal.ca}, \mail{Roman.Smirnov@dal.ca}}


\ArticleDates{Received ???, in final form ????; Published online ????}

\Abstract{ In this paper, we establish a rigorous geometric framework that applies the formal architecture of finite-dimensional bi-Hamiltonian structures and Nambu-Poisson mechanics to macroeconomic growth theory. Moving beyond static empirical correlations, we deploy Ryuzo Sato's principle of simultaneous holotheticity to demonstrate that economic production functions emerge natively as stable, time-independent geometric leaves of integrable phase flows. We construct a unified, hierarchical taxonomy of three fundamental economic growth regimes that sequentially generalize one another: the classical, unconstrained Cobb–Douglas model; the resource-limited, $S$-shaped (sigmoidal) ecosystem response; and the capacity-bounded ``overshoot-and-collapse" regime. Furthermore, we push this paradigm into non-smooth territory by investigating the structural boundary crises that occur when aggregate economic trajectories encounter definitive resource carrying capacity ceilings. We show that these capacity limits induce a catastrophic rank-collapse of the compatible contravariant Poisson pencil. By utilizing Filippov's convex reconstruction method and a subsequent Dirac bracket reduction, we successfully project the three-dimensional bi-Hamiltonian hierarchy onto a lower-dimensional constraint manifold. This geometric mechanism preserves the underlying algebraic duality and compatibility of the Poisson bivectors under discontinuous sliding regimes, providing a principled mathematical foundation for understanding structural locks, complexity ceilings, and systemic crises in economic systems.}


\Classification{37J06; 37J35; 37J39; 53D17; 70G65; 70H33; 91B55; 91B62} 

\begin{quote}
\hfill{\em “Nature imitates mathematics.”}
— Gian-Carlo Rota
\end{quote}

\section{Introduction}
\label{s1}

The formal architecture of bi-Hamiltonian mechanics represents one of the most profound conceptual evolutions in modern mathematical physics, standing as a monumental paradigm shift in our understanding of dynamical systems that is comparable in its fundamental importance to the discovery of 
Newton's second law, Lagrangian mechanics, and classical Hamiltonian systems. Far from being a rigid framework confined to conservative physical problems, the utility of bi-Hamiltonian structures has expanded well beyond the traditional boundaries of mathematical physics; notably, they have been successfully applied to complex bio-dynamical models, such as multi-dimensional generalizations of competitive Lotka–Volterra systems (see, for example, Damianou and Fernandes \cite{DF2002}), where compatible Poisson bivectors reveal hidden conservation laws and global stability properties in population dynamics.

The theory originated from what was initially perceived as an isolated algebraic curiosity: Andrew Lenard’s unpublished, accidental discovery of an infinite sequence of conserved quantities for the Korteweg–de Vries (KdV) equation (see \cite{PS2005} for historical details). The resulting Lenard recursion scheme was subsequently utilized by Miura {\em et al.} \cite{MGK1968} and Gardner {\em et al.} \cite{GGKM1974} as an explicit algorithmic tool to generate the infinite hierarchy of conserved quantities for the KdV equation. Prior to Franco Magri’s seminal 1978 paper \cite{FM1978}, this iterative scheme was viewed as an isolated computational anomaly unique to the KdV equation. 
Several authors of the pioneering papers prominently cited a ``private communication with Andrew Lenard''~\cite{MGK1968, GGKM1974}, crediting him with the singular spark that initiated the entire development. Israel Gel’fand and Irene Dorfman~\cite{GD1979} even explicitly invoked the term ``the Lenard scheme,'' noting that they first learned of it from a lecture delivered by Peter Lax. 

Around the same time, Franco Magri demonstrated that this purely algebraic mechanism could be elevated into a universal geometric framework applicable to a wide variety of nonlinear partial differential equations (PDEs), including the modified KdV (mKdV) equation, the nonlinear Schr\"{o}dinger (NLS) equation, and the Harry Dym equation. Magri's fundamental insight was that the Lenard recursion is the natural geometric manifestation of a manifold endowed with two compatible Poisson bivector --- a framework known today as the Lenard–Magri scheme or the bi-Hamiltonian formalism.  This compatibility generates a bi-Hamiltonian pencil, a structure that has revolutionized the classification and analysis of infinite-dimensional integrable systems. 

Shortly thereafter, Gel'fand and Dorfman \cite{GD1979} successfully extended this machinery to finite-dimensional dynamical systems. By formalizing bi-Hamiltonian structures using algebraic pairs of Poisson bivectors and multi-Hamiltonian formalisms, they bridged the gap between infinite-dimensional PDEs and finite-dimensional ordinary differential equations (ODEs). Crucially, they established that the compatibility of two Poisson bivectors was equivalent to the vanishing of the Nijenhuis tensor \cite{AN1951} of the corresponding recursion operator constructed out of them.
 
 This geometric framework was further refined and generalized by leading contributors to modern Poisson geometry. Notably, Rui Fernandes \cite{RLF1994} introduced deep insights into the classification, linearizability, and global behavior of bi-Hamiltonian systems, mapping out their core topological constraints and establishing a precise characterization of complete integrability in the Liouville–Arnold sense \cite{LJ1855, VIA1988} for a bi-Hamiltonian system. Following these discoveries, Oleg Bogoyavlenskij \cite{OB96, OB98} and others (see, for example, \cite{VMP2025, VFMP2025, RLF1993, BF1982, PL2002, PVV2021, MS2015, RGS1, RGS3, RGS6, RGS6A, TT2022} and the relevant references therein) advanced the field by developing a comprehensive theory of master symmetries, dynamically coupled but incompatible Poisson bivectors, and the explicit integration of finite-dimensional hydrodynamic and physical systems via multi-Hamiltonian hierarchies.

The primary objective of this paper is to demonstrate that the bi-Hamiltonian framework --- traditionally reserved mostly for physical systems --- can be naturally and fruitfully applied to fundamental problems in economic growth theory. Historically, macroeconomic modeling has relied on static, empirical functional forms to represent production capacities, often leaving the microeconomic and dynamic foundations underdetermined. By reinterpreting economic growth through the lens of differential geometry, we show that the derivation of production functions and the rigorous analysis of their structural properties can be treated as an intrinsic problem of integrable phase flows. This work serves as a natural and direct extension of an ongoing research program \cite{SW2020, SW2019, SW2021, SWW2022, SW2024, S2025, S2026, SV2026} initiated in \cite{SW2020}. In our previous collaborative papers, we established the utility of the bi-Hamiltonian approach for generating valid economic production functions from underlying Lie group symmetries and evolutionary vector fields \cite{SW2019, SW2021}. In this manuscript, prepared for the special session of SIGMA dedicated to the 60th anniversary of Rui Fernandes, we push this paradigm into non-smooth territory. We utilize the bi-Hamiltonian formulation to deconstruct a generalized, non-separable asset-allocation model. We show not only how standard macro-dynamics emerge as Hamiltonian invariants, but also how these dual Poisson structures behave under the discontinuous, non-smooth conditions of a Filippov sliding regime. In doing so, we uncover a novel geometric mechanism: the collapse of a 3D bi-Hamiltonian hierarchy onto a lower-dimensional constraint manifold, providing a rigorous structural foundation for economic complexity ceilings.

This paper is organized as follows. In Section~\ref{s2}, we review the foundational 
geometric theory of finite-dimensional bi-Hamiltonian pencils rooted in 
the  compatibility criteria for Poisson bivectors and the conditions 
for complete integrability in the sense of Liouville--Arnol'd. Section~\ref{s3} bridges 
this mathematical architecture with macroeconomic theory, tracking the evolution 
of production functions from historical empirical heuristics to continuous Lie group 
invariants governed by simultaneous holotheticity. In Section~\ref{s4}, we specialize the 
bi-Hamiltonian formalism to three-dimensional spaces, establishing the geometric 
duality between Nambu-Poisson mechanics contractions and compatible, singular Poisson pairs. 

This geometric taxonomy is then explicitly deployed across three hierarchical economic 
growth regimes that sequentially generalize one another. Section~\ref{s41} analyzes Case~I, 
the unconstrained Cobb--Douglas framework, utilizing a scale-invariant log-canonical 
volume form to endogenously generate a production function enjoying constant returns to scale. Section~\ref{s42} expands 
this framework to Case~II, a resource-limited, $S$-shaped (sigmoidal) ecosystem response 
characterized by a non-local sigmoidal volume form. Section~\ref{s43} investigates Case~III, 
the capacity-bounded ``overshoot-and-collapse'' regime, wherein the smooth 
bi-Hamiltonian integrability breaks down at structural resource boundaries; we resolve 
this systemic crisis by projecting the three-dimensional dynamics onto a 
lower-dimensional constraint manifold via a Filippov sliding reconstruction and a 
subsequent Dirac bracket reduction. Finally, Section~\ref{s5} presents our conclusions and 
proposes avenues for future geometric and data-driven macroeconomic research.

\section{Geometric Foundations of Finite-Dimensional Bi-Hamiltonian Pencils  }
\label{s2}


Let $X_H$ be a Hamiltonian vector field defined on a Poisson manifold $(M, \pi)$. To characterize the evolution of $X_H$  as an integrable geometric phase flow, we utilize the framework of Poisson calculus. Recall, a Poisson bivector $\pi$ on $M$ is uniquely defined by a smooth, skew-symmetric contravariant bivector field $\pi \in \mathcal{T}^{(2,0)}(M)$ that satisfies the intrinsic integrability criterion:
\begin{equation}
[\pi, \pi] = 0,
\label{eq:poisson_integrability}
\end{equation} 
where $[\cdot, \cdot]$ denotes the canonical Schouten--Nijenhuis bracket \cite{AN1955-1, AN1955-2} mapping skew-symmetric tensors, a special case of the general Schouten bracket introduced in \cite{JS1940} (see \cite{K-S2021} for more details and references). This geometric condition is algebraically equivalent to enforcing the Jacobi identity for the corresponding Poisson bracket of smooth functions, defined via the differential contraction $$\{f, g\}:= \pi(\mbox{d}f, \mbox{d}g) = [[\pi, f], g]$$ for all $f, g \in C^\infty(M)$. Given a choice of a smooth economic Hamiltonian function $H \in C^\infty(M)$, the corresponding autonomous phase velocity is generated by contracting the differential of $H$ with the Poisson bivector to yield the vector field $X_H$ given by
\begin{equation}
\label{hamvf}    
X_H = \pi(\mbox{d}H, \cdot) = [\pi, H].
\end{equation}
Importantly, if the Poisson bivector $\pi$ in (\ref{hamvf}) is non-degenerate (i.e., $\pi^{-1} = \omega$ is a symplectic form), the Hamiltonian vector field $X_H$ can be recovered from $H$ via the formula (\ref{hamvf}) and defined globally. The situation is different when $\pi$ is degenerate --- see Section \ref{s4} for more details. 

Following the foundational formulations of Magri \cite{FM1978} and Gel'fand $\&$ Dorfman \cite{GD1979}, an evolutionary growth vector field $X_{H_1, H_2}$ is said to be bi-Hamiltonian if it can be simultaneously represented as a Hamiltonian system with respect to two distinct,  independent Poisson bivectors $\pi_1$ and $\pi_2$:
\begin{equation}
X_{H_1, H_2} = [\pi_1, H_1] = [\pi_2, H_2].
\label{eq:bihamiltonian_pencil}
\end{equation}
To ensure the complete integrability of the underlying bi-Hamiltonian flow, these dual geometric structures must be compatible. This dictates that any linear combination within the pencil $\pi_1 + \lambda \pi_2$ must remain a valid Poisson bivector for any arbitrary scalar $\lambda \in \mathbb{R}$. Under the Schouten--Nijenhuis calculus, this compatibility restriction simplifies elegantly to the single bilinear vanishing condition:
\begin{equation}
[\pi_1, \pi_2] = 0.
\label{eq:poisson_compatibility}
\end{equation}

Assuming $\pi_1$ is non-degenerate on an open dense subset of the state space, this Poisson pair naturally induces a mixed $(1,1)$-tensor field acting as a recursion operator \cite{Olver1977}, defined by the product $A:= \pi_2 \pi_1^{-1} \in \mathcal{T}^{(1,1)}(M)$. As proven by Gel'fand and Dorfman \cite{GD1979}, the compatibility requirement \eqref{eq:poisson_compatibility} is structurally equivalent to the vanishing of the eponymous $(1,2)$-Nijenhuis tensor $N_A \in \mathcal{T}^{(1,2)}(M)$ \cite{AN1951}, constructed coordinate-free for any vector fields $X, Y \in \mathfrak{X}(M)$ as:
\begin{equation}
N_A(X, Y) = A^2[X, Y] + [AX, AY] - A\big([AX, Y] + [X, AY]\big).
\label{eq:nijenhuis_tensor}
\end{equation}
Recall, the Nijenhuis tensor (\ref{eq:nijenhuis_tensor}) vanishes identically if and only if the eigenvectors of $A$ form integrable codimension-one distributions on an $n$-dimensional manifold $M$ (hypersurfaces, historically denoted by $X_{n-1}$). These integrable hypersurface distributions --- generated by the eigenvalues and eigenvectors of Killing two-tensors in constant curvature spaces — have been studied extensively (see, e.g.,~\cite{CMS2011, CMS2017, HMS2005, HMS2009, MST2002, MST2004}). However, the Nijenhuis tensor cannot be applied directly to investigate such distributions~\cite{BMS2001}: imposing the Killing‑tensor equation and the vanishing of the Nijenhuis tensor simultaneously gives overly restrictive conditions. Several recipes are available in this context. For example, Sergio Benenti \cite{SB1997} to verify the integrability of the eigenvectors of a Killing two-tensor $\hat{K} = Kg^{-1}$ defined an auxiliary conformal Killing tensor $G$ with the same eigenvectors and then applied the formula (\ref{eq:nijenhuis_tensor}) to $\hat{G} = Gg^{-1}$ to verify that the eigenvectors of the original Killing tensor were surface-forming. Alternatively, one can apply the {\em Tonolo-Schouten-Nijenhuis (TSN)} condition (see, for example, p. 687 in \cite{HMS2005}) or the {\em Haantjes tensor} \cite{H1955}, which is a natural generalization of the Nijenhuis tensor \cite{AN1951}.  In our context, the identity $N_A = 0$ guarantees that the eigenvalues of the recursion operator $A$ form a mutually commuting set of functionally independent first integrals that under some additional algebraic and dimensional conditions afford complete integrability of $X_{H_1, H_2}$ in the Liouville--Arnold sense \cite{LJ1855, VIA1988} --- see Magri and Morosi \cite{MM1984} for requisite details. These two results \cite{GD1979, MM1984}  form a foundation for a construction that guarantees complete integrability of a bi-Hamiltonian vector field $X_{H_1, H_2}$ defined by (\ref{eq:bihamiltonian_pencil}) on an even-dimensional manifold $M$ with respect to two Poisson bivectors $\pi_1$ and $\pi_2$ at least one of which is non-degenerate --- see, for example,  \cite{RGS6A} for more details. 

In his seminal 1994 paper titled {\em ``Completely integrable bi-Hamiltonian systems"} \cite{RLF1994},  Fernandes established a deep, intrinsic geometric link between the Liouville-Arnol'd complete integrability of a Hamiltonian system and the existence of a bi-Hamiltonian structure. Specifically, he proved that under mild, natural hypotheses, a compatible second non-degenerate Poisson bivector exists in a neighborhood of the invariant torus such that the recursion operator exhibits a maximal set of functionally independent eigenvalues if and only if the graph of the Hamiltonian function \(H(I)\) is a hypersurface of translation with respect to the affine structure generated by the action coordinates. This result shows that the existence of a bi-Hamiltonian structure defined by two non-degenerate Poisson bivector is a very rigid construction.  While Fernandes proved exactly when it is possible, subsequent research showed that the set of functions that are additively separable after an affine transformation is topologically meagre (very sparse) --- see a recent paper by Boualem and Brouzet \cite{BB2021}. Of course, not all integrable Hamiltonian systems fall into this category: for example, isochronous systems (such as the unperturbed Kepler problem)  do not satisfy the Kolmogorov non-degeneracy condition $\left(\frac{\partial^2 H}{\partial I_i \partial I_i}\right) \not=0$ required by Fernandes' characterization. 

In view of Fernandes’ theorem, which severely limits standard bi-Hamiltonian structures to highly restrictive hypersurfaces of translation,  Bogoyavlenskij introduced the concept of dynamically compatible but algebraically incompatible Poisson bivectors. Recognizing that the compatibility condition (\ref{eq:poisson_compatibility}) was too rigid to describe generic Liouville-Arnol'd integrable systems, Bogoyavlenskij \cite{OB96} shifted focus to Poisson bivectors that are simply invariant under the system's phase flow. By decoupling integrability from algebraic compatibility, his framework bypassed Fernandes' strict additive separability roadblocks. This allowed him to prove that master hierarchies of constants of motion still naturally emerge from incompatible pairs, providing a robust geometric foundation for the vast majority of physical systems that fail Fernandes’ criteria.

In the subsequent sections, this precise bi-Hamiltonian framework will be deployed to analyze scale-invariant vector fields within economic growth theory. Specifically, we will construct a concrete 3D Poisson architecture where macroeconomic production functions emerge naturally as stable geometric leaves of a bi-Hamiltonian system. Then we will investigate the structural limits of this formalism when boundary constraints introduce non-smoothness. We demonstrate that resource ceilings induce a catastrophic rank-collapse of the compatible Poisson pencil, a geometric crisis that we successfully resolve by applying Filippov’s convex reconstruction to project the broken dynamics onto an integrated Hamiltonian flow. 

\section{Production Functions in Economic Growth Theory: From Empirical Models to Geometric Invariants}
\label{s3} 

In what follows, we focus on the concept of a \emph{production function} in economic growth theory. In this context, a production function expresses physical output as a function of various input factors, such as capital, labor, land, and energy. Most production functions studied to date possess the following functional form: 
\begin{equation}
\label{pf}
Y = F(K, L),
\end{equation}
where the production output $Y$ is determined by capital $K$ and labor $L$. 

Production functions that are homogeneous of degree one are of paramount importance in economics. Such functions satisfy the relation
\begin{equation}
\label{crts}
\lambda Y = F(\lambda K, \lambda L) \quad \text{for all } \lambda > 0,
\end{equation}
and are said to exhibit \emph{constant returns to scale (CRTS)}. Under~\eqref{crts}, scaling both inputs $K$ and $L$ by a factor of $\lambda$ yields an identical scaling of the output $Y$. 

From a mathematical perspective, this homogeneity permits the introduction of the projective (or per capita) coordinates 
\begin{equation}
\label{pc}
x = \frac{K}{L} \quad \text{and} \quad y = \frac{Y}{L},
\end{equation}
which represent capital per capita and production per capita, respectively. By setting $\lambda = 1/L$, the functional form~\eqref{pf} reduces to 
\begin{equation}
\frac{Y}{L} = F\left(\frac{K}{L}, 1\right),
\end{equation}
or, expressed in terms of the coordinates $(x,y)$: 
\begin{equation}
\label{ppf}
y = f(x), 
\end{equation}
where $f(x) \coloneqq F(x, 1)$.

Historically, the inception, development, and formalization of production functions within macroeconomic modeling have evolved across distinct speculative and foundational paradigms. The trajectory maps a continuous transition from ad hoc mathematical formulations to rigorous data-driven geometric invariants, intimately intertwined with the validation of functional income distribution laws.

\begin{enumerate}
    \item \textbf{The Mathematical Convenience Phase:} In early macroeconomic formulations, functional forms mapping capital ($K$) and labor ($L$) inputs to aggregate output ($Y$) were introduced strictly \textit{ad hoc}. These primitive models lacked empirical or systemic microfoundations, serving merely as mathematically convenient, continuous idealizations to analyze economic performance, marginal productivities, and returns to scale. To model the production output, the economists of the later part of the 19th century -- earlier part of the 20th century employed the functional form 
    \begin{equation} 
    \label{CD}
    Y = A K^{\alpha} L^{1-\alpha}
    \end{equation}
that was  introduced {\em ad hoc}  and studied, for example,  by Knut Wicksell, Philip Wicksteed, and L\'{e}on Walras (see
Humphrey \cite{H1997} for more details and references). For this reason, the production function given by (\ref{CD}) is still called by some authors the {\em Wicksellian production function} \cite{RWG2026}. We note that the function (\ref{CD}) is CRTS and in terms of the projective coordinates (\ref{pc}) it assumes the following simple form:
\begin{equation}
\label{pCD}
y = A x^{\alpha},
\end{equation}
It must be noted that within the Solow-Swan framework \cite{RS1956, TS1956} and its generalizations, production is normally modeled via a Cobb-Douglas function in per-capita (projective) coordinates (\ref{pCD}), subject to an exogenous rate of technological change. Consequently, the shift factors within the functional form (\ref{CD}) or (\ref{pCD}) are treated as parameters given from outside the model. 
    \item \textbf{The Empirical Modeling Phase and Functional Distribution:} This paradigm was fundamentally redefined by the seminal work of Charles Cobb and Paul Douglas in their celebrated 1928 paper \cite{CD1928}. Moving away from arbitrary mathematical constructs, they introduced an analytically stable formulation that could be fitted directly to empirical historical data from the US manufacturing sector spanning 1899--1922. Crucially, the resulting Cobb--Douglas production function, given by (\ref{CD}) for $A = 1.01$ (representing technical progress) and $\alpha = 0.25$, was validated by its alignment with the evolution of the American economy during this period. Over the subsequent two decades, Paul Douglas and a team of statisticians successfully tested this functional form against various other datasets, further justifying its empirical relevance --- see \cite{PD1976} for comprehensive details and references. This extensive testing effectively ``legitimized| the Cobb--Douglas specification by anchoring it to empirical reality. However, this framework left critical questions unanswered. For instance, given a specific economic dataset --- such as the one originally studied in \cite{CD1928} --- are there other functional forms or variations that fit the data equally well? The answer to this question is affirmative \cite{SWW2022, SW2024}, implying that merely fitting the traditional functional form (\ref{CD}) to a dataset does not capture all the essential information required to fully understand the dynamic evolution of the variables $K = K(t)$, $L = L(t)$, and $Y = Y(t)$.

    \item \textbf{The Geometric and Lie Group Invariant Phase:} A profound conceptual shift occurred when the renowned Japanese economist Ryuzo Sato introduced the formal machinery of continuous Lie groups into macroeconomic growth theory \cite{RS1981} (see also Sato and Ramachadran \cite{SR2014}). Sato demonstrated that production functions should not be treated as static empirical correlations. Instead, utilizing his newly developed geometric framework of \textit{simultaneous holotheticity}, he proved that these functions emerge as multi-parameter Lie group invariants of an underlying distribution of vector fields representing exponential technological change. This approach elevated production functions to the status of fundamental geometric constants of motion. 

Within the context of Lie group theory applied to economic growth models, simultaneous holotheticity describes a multi-sector economy where distinct sectors operate under the same aggregate production function while experiencing non-uniform rates of technical progress. At the level of infinitesimal actions, this economic assumption mandates that the aggregate production function $\phi(K, L, Y) = 0$ emerge as a mutual, time-independent invariant of a multi-dimensional distribution of vector fields representing simultaneous scaling transformations across sectors. By leveraging this mathematical apparatus, Sato successfully derived the production function (\ref{CD}) as a time-independent invariant determined by an integrable distribution of two vector fields that define the exponential growth of the functions $K = K(t)$, $L = L(t)$, and $Y = Y(t)$.

\item \textbf{Data-Driven Dynamical Systems Phase:} In the modern data-driven context pioneered by Smirnov and Wang \cite{SW2024}, which utilizes and combines the two previous approaches, this multi-parameter framework simplifies elegantly into a unified one-parameter Lie (semi-)group action. Rather than artificially imposing simultaneous holotheticity across disjoint sectors, the underlying economic velocity vector field is extracted directly from the empirical trajectories of the aggregate phase-space variables $(K, L, Y)$, expressed as time series determined by observed data. For instance, when growth is characterized by robust exponential expansion in $K = K(t)$, $L = L(t)$, and $Y = Y(t)$, the corresponding infinitesimal generator induces a continuous one-parameter (semi-)group transformation. The family of Cobb–Douglas production functions is then recovered with strict geometric rigor, emerging naturally as the stable, time-independent invariant surfaces under this one-parameter (semi-)group flow, if and only if the output elasticities satisfy the linear orthogonality constraint dictated by the system's growth rates --- see \cite{SWW2022, SW2024} for further details.

Once an integrable distribution of two vector fields is established via Sato's simultaneous holotheticity condition --- where the first field defines the evolution of the economic trajectories $K = K(t)$, $L = L(t)$, and $Y = Y(t)$, while the second guarantees homogeneity of degree one (\ref{crts}) --- advanced symmetry methods \cite{Olver86} can be applied. Specifically, the infinitesimal generator of economic growth is extended to its first prolongation $\text{Pr}^{(1)}(\mathbf{u})$ on the first-order jet bundle $J^1\pi$, which is equipped with contact coordinates $(K, L, Y, Y_K, Y_L)$ where $Y_K = \partial Y/\partial K$ and $Y_L = \partial Y/\partial L$. This framework enables the derivation of differential invariants, including those explicitly dependent on the marginal productivities $Y_K$ and $Y_L$. Notably, it has been demonstrated with strict geometric rigor that the validity of Bowley's Law is an intrinsic characterization unique to exponential growth \cite{SW2020, SV2026}. When the economy undergoes robust exponential expansion, the classical wage and capital shares emerge naturally as time-independent differential invariants of the prolonged group action.

Concurrently, this framework exposes the structural limitations of classical models. When unconstrained exponential growth is generalized to a realistic, capacity-limited logistic growth model, the underlying diffeomorphism breaks the scale invariance. On the resulting logistic production manifold, the labor share $s_L$ becomes explicitly time-dependent, mathematically elucidating the contemporary breakdown of Bowley's Law observed in post-1980 empirical data \cite{SW2020}.
\end{enumerate}

Consequently, the production function is structurally vindicated within modern geometry not as an empirical heuristic, but as a stable geometric leaf or invariant foliation arising from the (data-driven) dynamics of the state space.

\begin{example}
Indeed, consider the logistic model introduced in \cite{SW2020} (see also \cite{SW2024}) and  given by 
\begin{equation}
\label{logistic} 
\dot{K} = b_1K\left(1 - \frac{K}{N_K}\right), \,
\dot{L} = b_2L\left(1 - \frac{L}{N_L}\right), \, 
\dot{Y} = b_3Y\left(1 - \frac{Y}{N_Y}\right), 
\end{equation}
where $b_1$, $b_2$, and $b_3$ are the growth rates for capital, labor, and production, while $N_K$, $N_L$, and $N_Y$ are the corresponding carrying capacities. It is possible to show that the production function
\begin{equation}
Y = \frac{N_Y K^{\alpha}L^{\beta}}{K^{\alpha}L^{\beta} + \Omega|N_K-K|^{\alpha}|N_L-L|^{\beta}}, 
\label{prod2}
\end{equation}
where $\alpha  = - a_1/a_3$, $\beta = -a_2/a_3$ and $\Omega = \left(\frac{N_Y  - Y_0}{Y_0}\right)\left(\frac{K_0}{|N_K - K_0|}\right)^{\alpha}\left(\frac{L_0}{|N_L-L_0|}\right)^{\beta},$
is a time-independent invariant of the flow generated by (\ref{logistic}), provided 
${\bf a}\cdot{\bf b} = a_1b_1 + a_2b_2 + a_3b_3 = 0, $
where the components of the vector ${\bf b} = <b_1, b_2, b_3>$ are the growth rates of capital, labor, and production respectively defined by (\ref{logistic}), while the components of the vector ${\bf a} = <a_1, a_2, a_3>$ are arbitrary parameters constrained by the condition ${\bf a} \not= {\bf 0}$, following from $a_3 \not=0$. The parameters $b_1$, $b_2$, $b_3$, $N_K$, $N_L$, and $N_Y$ can be extracted from time-series representing growth in $K$, $L$, and $Y$ that is (close to) logistic.

\end{example}

\section{Bi-Hamiltonian Systems in 3D Spaces: Economic Aspects}
\label{s4}

We first note that the existence of a bi-Hamiltonian structure (\ref{eq:bihamiltonian_pencil}) for a vector field \(X \in \mathfrak{X}(M)\) on a three-dimensional manifold \(M\) (\(\dim M = 3\)) is a very strong condition that implies that the dynamical system defined by \(X\) is completely integrable. This means the two Hamiltonian functions \(H_{1}\) and \(H_{2}\) (provided they are functionally independent, meaning their gradients are linearly independent across an open and dense subset of the space) serve as first integrals that define the 1D trajectories as intersections of the 2D hypersurfaces defined by \(H_1(K, L, Y) = c_1\) and \(H_2(K, L, Y) = c_2\), respectively. Once the trajectories are confined to 1D curves formed by their intersections, the problem of solving the differential equations defined by \(X\) reduces to a single integration (a quadrature) along that curve. 

In what follows, we will investigate the dynamical systems defined within the framework of the economic growth theory that are defined by the evolution of three economic quantities as functions of time, namely, $K = K(t)$ (capital), $L = L(t)$ (labor), and $Y = Y(t)$ (production) in $M= \mathbb{R}_+^3$. What is the rationale for seeking the bi-Hamiltonian structure (\ref{eq:bihamiltonian_pencil}) for a 3D dynamical system apart from the presence of two independent invariants ($H_1$ and $H_2$)? 

For one, a bi-Hamiltonian structure, even in a 3D space, yields a rich structure of conserved quantities preserved by the bi-Hamiltonian vector field (see, for example, Vizarella {\em et. al} \cite{VMP2025} and the relevant references therein). 

Next, as we shall see below, the bi-Hamiltonian structure provides the exact algebraic mechanism required to handle non-smooth structural boundaries of the Filippov sliding regime. When the trajectory reaches the capacity line of the switching manifold $\Sigma$, the gradient vector field experiences an infinite vertical jump discontinuity, causing the first Poisson bivector $(\pi_1)$ to encounter a local blow-up singularity.  Because the framework provides a second Poisson bivector ($\pi_2$), we can exploit the property that the constraint function acts as a Casimir invariant of $\pi_2$.  This mathematical duality allows for a clean projection of the 3D dynamics onto the codimension-one switching surface via a Dirac bracket reduction ($\pi_{Dirac}$) without introducing mathematical infinities or breaking the underlying geometric structure.   

Moreover, a vector field admitting the bi-Hamiltonian structure defined by compatible two Poisson bivectors $\pi_1$ and $\pi_2$ satisfying (\ref{eq:poisson_compatibility}) exposes the underlying foliation topology of the economic phase space defined by $K$, $L$, and $Y$. By invoking the Frobenius theorem, the 3D state space is cleanly sliced into a regular foliation of two-dimensional integral submanifolds parametrized precisely by the constant capital-labor ratio (see below).  The bi-Hamiltonian formulation gives us the analytical tools --- such as normal projection vectors and wedge products --- to explicitly prove where the standard smooth transversality of this foliation breaks down. This allows us to formalize the precise point of contact where a 2D smooth leaf collapses into 1D constrained path, forcing a structural lock between independent economic variables.

Last, but not least, from a computational perspective, identifying the explicit Poisson bivectors and the scalar Jacobi multiplier ($\rho$) is essential for designing structure-preserving numerical simulations. Traditional numerical integrators (e.g., the standard Runge-Kutta schemes) suffer from artificial energy dissipation or accumulation, causing trajectories to rapidly drift away from the invariant manifolds over extended timelines. By utilizing the explicit bi-Hamiltonian structure (\ref{eq:bihamiltonian_pencil}), one can deploy specialized energy-preserving field (AVF) or Kahan splitting algorithms. These geometric integrators preserve the phase space volume and force the numerical trajectories to remain strictly bound to the true structural manifolds of the system \cite{CGMMOOQ2013, CMOQ2018, KU2026, U2021}. 

Define on $\mathfrak{X}(\mathbb{R}_+^3)$ the following scaling infinitesimal symmetry $X_s$ given by 
\begin{equation}
\label{sym}
X_s =  K\frac{\partial}{\partial K} + L\frac{\partial}{\partial L} + Y\frac{\partial}{\partial Y}. 
\end{equation}
It was proven in \cite{SV2026} that the most general vector field $X \in \mathfrak{X}(\mathbb{R}_+^3)$, satisfying the condition $[X_s, X]  = 0$ is given by 
\begin{equation}
\label{genX}
X = L \cdot \Phi_1(x,y)\frac{\partial }{\partial K} + L\cdot \Phi_2(x,y)\frac{\partial}{\partial L} + L\cdot  \Phi_3(x,y)\frac{\partial}{\partial Y},
\end{equation}
where $\Phi_i(x,y)$, $i=1,2,3$ are arbitrary functions of $x = K/L$, $y= Y/L$. 
Clearly, in terms of the projective coordinates $x = K/L$ and $y  = Y/L$ the vector field $X$ given by (\ref{genX}) reduces to 
\begin{equation}
\label{gen2X}
\widetilde{X}  =\left[\Phi_1(x, y) - x\Phi_2(x, y)\right]\frac{\partial}{\partial x} + \left[\Phi_3(x, y) - y\Phi_2(x, y)\right]\frac{\partial}{\partial y}. 
\end{equation}

From a geometric perspective, Sato's idea of simultaneous holotheticity was realized in \cite{SV2026} within the framework of the following

\begin{definition}
\label{d1}
A production function $Y = F(K, L)$ is said to be {\em simultaneously holothetic} with respect to a scale dilation group $G_S$ (generated by the vector field $X_s$ given by (\ref{sym})) and a technological or evolutionary transformation group $G_X$ (generated by a vector field $X$) if the production hypersurface ${\cal M} = \{(K, L, Y) \in \mathbb{R}_+^3|Y- F(K, L) = 0\}$ is a mutually invariant locus under the action of both groups. 
\end{definition}
Next, the following theorem was proven in \cite{SV2026}: 
\begin{theorem}[Holothetic Duality and Leaf Alignment]
\label{t1}
Let $X$ be the general scale-invariant vector field on the state space $M = \mathbb{R}_+^3$ satisfying $[X_s, X] = 0$, given by (\ref{genX}), where $X_s$ is the scaling symmetry (\ref{sym}). 
Then, the production function $Y = F(K, L)$ exhibits constant returns to scale and satisfies $X(Y - F(K, L)) = 0$ if and only if the hypersurface ${\cal M}$ is a 2-dimensional integral leaf of the involutive distribution ${\cal D} = \mbox{span}\{X_s, X\}$.
\end{theorem}
\begin{example}
The vector field $X_{\ell}$ determined by (\ref{logistic}) and given by 
\begin{equation}
 \label{logvf}   
X_{\ell} =  b_1K\left(1 - \frac{K}{N_K}\right)\frac{\partial}{\partial K} + b_2L\left(1 - \frac{L}{N_L}\right)\frac{\partial}{\partial L} + b_3Y\left(1 - \frac{Y}{N_Y}\right)\frac{\partial}{\partial Y}
\end{equation}
is not  of the form (\ref{genX}). For this reason, it does not commute with the scaling symmetry (\ref{sym}) and the production function (\ref{prod2}), in view of Theorem \ref{t1},  does not exhibit CRTS. 
\end{example}

To rigorously formulate a Hamiltonian system within a three-dimensional configuration space like \(\mathbb{R}_{+}^{3}\), one must bypass the standard even-dimensional symplectic trick and instead deploy the framework of Nambu-Poisson mechanics. Because any \(3 \times 3\) skew-symmetric tensor is structurally degenerate, a unique phase velocity cannot be recovered by inverting a single Hamiltonian function. Instead, a valid 3D system requires a choice of a top-degree volume form \(\Omega \) and a pair of functionally independent conserved quantities (\(H_1, H_2\)). The corresponding autonomous vector field is then uniquely generated via the dual Nambu contraction \(\iota_X \Omega = \mbox{d}H_1 \wedge \mbox{d}H_2\), which yields the classical cross-product flow lines \(X = \Omega^{-1}(\mbox{d}H_1 \wedge \mbox{d}H_2)\) mapped along the intersecting level surfaces of the two invariants. This architecture naturally uncovers an elegant geometric duality: by cross-contracting each individual invariant against the background volume form, one constructs a pair of compatible Poisson bivectors, \(\pi_1 = \Omega^{-1}(\mbox{d}H_2 \wedge \cdot)\) and \mbox{\(\pi_2 = \Omega^{-1}(\mbox{d}H_1 \wedge \cdot)\)}, which allows the identical vector field to be represented as a bi-Hamiltonian system satisfying (\ref{eq:bihamiltonian_pencil}),  where $X = X_{H_1, H_2}$. Under this setup, the functions dynamically exchange roles: the Hamiltonian \(H_{1}\) of the first structure acts identically as the structural Casimir invariant spanning the kernel of the second structure $ [\pi_2, H_1] \equiv {\bf 0}$, while \(H_{2}\) acts as the Casimir of the first  $[\pi_1, H_2] \equiv {\bf 0}$. 

Crucially, such Hamiltonian fields do not automatically preserve the background phase volume. Recall, on a more general Poisson manifold (like the 3D space in this paper), a Hamiltonian vector field does not automatically preserve an arbitrary or standard Euclidean volume form. It will only preserve a given volume form if the underlying Poisson bivector is unimodular, meaning its {\em modular vector field}  vanishes with respect to that volume form. In an odd-dimensional Poisson framework, a field is volume-preserving if and only if the structure is unimodular with respect to \(\Omega \) (the volume form). If the global modular vector field \(\phi _{\Omega }\) does not vanish, the flow undergoes a localized divergence --- see Crainic {\em et al.} \cite{CFM2021} for more details and proofs.

Now, let us investigate the vector field (\ref{genX}) from this viewpoint. Indeed, we first observe that, of course, not every vector field given by (\ref{genX}) is in fact a volume-preserving vector field. Under that scenario, the structure is unimodular (divergence-free with respect to the standard Euclidean volume form $\Omega = \mbox{d}K \wedge \mbox{d}L \wedge \mbox{d} Y$). Next,  we calculate $\text{div}_\Omega(X)=\nabla \cdot X$: 
$$\text{div}_\Omega(X) = \frac{\partial }{\partial K} (L\Phi_1) + \frac{\partial }{\partial L}(L\Phi_2) + \frac{\partial }{\partial Y}(L\Phi_3)$$
Applying the chain rule with respect to the projective coordinates $(x,y)$, the 3D divergence simplifies to: 
\begin{equation}
\label{ham1}
\text{div}_\Omega(X)= \frac{\partial \Phi_1}{\partial x} + \frac{\partial \Phi_3}{\partial y} + \Phi_2 - x\frac{\partial \Phi_2}{\partial x} - y \frac{\partial \Phi_2}{\partial y}. 
\end{equation}

The same condition ($\nabla \cdot \widetilde{X} = 0$) for the vector field (\ref{gen2X}) yields: 
\begin{equation}
\label{ham2}
\text{div}_\Omega(\widetilde{X})= \frac{\partial \Phi_1}{\partial x} + \frac{\partial \Phi_3}{\partial y} -2 \Phi_2 - x\frac{\partial \Phi_2}{\partial x} - y \frac{\partial \Phi_2}{\partial y}. 
\end{equation}
This discrepancy between the two calculations stems from a fundamental geometric principle in dynamical systems: divergence is not invariant under coordinate projections unless you adjust for the change in the underlying volume. Therefore, we conclude that the first step in implementing the procedure outlined above should be choosing the right volume form that can ``flatten" coordinates or make the vector field in question unimodular. 

\section{Case I: The Cobb-Douglas Regime}
\label{s41}

Consider Sato's simultaneous holotheticity condition afforded by the following integrable distribution: 
\begin{equation}
\label{D1}
 \mathcal{D}_1 =\left\{X_s= K\frac{\partial}{\partial K} + L\frac{\partial}{\partial L} + Y\frac{\partial}{\partial Y}, \,  X_1= b_1K\frac{\partial}{\partial K} + b_2L\frac{\partial}{\partial L} + b_3Y\frac{\partial}{\partial Y} \right\}.
 \end{equation}
We note that the vector field $X_1$ is a limiting case of (\ref{logvf}) when $N_K, N_L, N_Y \to \infty$. 
Clearly, $X_1$ is of the form (\ref{genX}) for $\Phi_1 = b_1 x$, $\Phi_2 = b_2$, and $\Phi_3 = b_3y$. Next, in view of Theorem \ref{t1}, we conclude that the underlying production function arising as a time-independent invariant of $\mathcal{D}_1$ is of the CRST type. 

We first analyze the linear scaling vector field \(X_1 = b_1 K \frac{\partial}{\partial K} + b_2 L \frac{\partial}{\partial L} + b_3 Y \frac{\partial}{\partial Y}\) under the standard flat Euclidean volume form:
\begin{equation}
\label{eV}
\Omega _{\text{flat}}=\mbox{d}K\land \mbox{d}L\land \mbox{d}Y.
\end{equation}
Computing the standard 3D divergence, using either the formula (\ref{ham1}) or (\ref{ham2}), yields: $$\text{div}_{\Omega _{\text{flat}}}(X_{1})=\frac{\partial (b_{1}K)}{\partial K}+\frac{\partial (b_{2}L)}{\partial L}+\frac{\partial (b_{3}Y)}{\partial Y}=b_{1}+b_{2}+b_{3}.$$ 

For \(X_{1}\) to be volume-preserving (unimodular) under flat space, we should require \(\text{div}_{\Omega_{\text{flat}}}(X_1) = 0\), which mandates the parametric constraint \(b_1 + b_2 + b_3 = 0\). From a macroeconomic standpoint, this condition is entirely unrealistic, as sustainable economic growth requires strictly positive parameters (\(b_i > 0\)).

To reconcile our dynamical system with economic reality, we must abandon flat space and equip the orthant \(\mathbb{R}_{+}^{3}\) with a volume form that respects scale invariance. We apply a conformal gauge transformation to construct the log-canonical volume form:
\begin{equation}
\label{logV}
\Omega _{\log }=\frac{1}{K \cdot L \cdot Y}\mbox{d}K\land \mbox{d}L\land \mbox{d}Y.
\end{equation}
Re-evaluating the generalized  divergence under \(\Omega _{\log }\) yields:
$$
\text{div}_{\Omega _{\log }}(X_{1})=KLY\left[\frac{\partial }{\partial K}\left(\frac{b_{1}}{LY}\right)+\frac{\partial }{\partial L}\left(\frac{b_{2}}{KY}\right)+\frac{\partial }{\partial Y}\left(\frac{b_{3}}{KL}\right)\right]\equiv 0. 
$$
This identity vanishes identically for any configuration of parameters. Under the log-canonical framework, the system becomes perfectly volume-preserving and unimodular for strictly positive growth rates (\(b_i > 0\)).

Because the configuration space is now non-local and volume-preserving, we deploy the framework of Nambu-Poisson mechanics under \(\Omega _{\log }\) to extract the absolute invariants. The autonomous flow lines are generated via the dual Nambu contraction 
\begin{equation}
\label{of}
\iota_{X_1}\Omega_{\log} = \frac{b_1b_2}{b_3}\mbox{d}H_1 \wedge \mbox{d}H_2,
\end{equation}
which yields the uncoupled log-canonical Hamiltonians:

$$ H_{1}=\ln (Y)-\frac{b_{3}}{b_{1}}\ln (K)\quad \text{and}\quad H_{2}=\ln (Y)-\frac{b_{3}}{b_{2}}\ln (L).$$

While the log-canonical Hamiltonians \(H_{1}\) and \(H_{2}\) constitute a complete basis of functionally independent invariants for the growth velocity field \(X_{1}\), they are not individually annihilated by the scaling symmetry field, since \(X_s(H_1) = 1 - \frac{b_3}{b_1}\) and \(X_s(H_2) = 1 - \frac{b_3}{b_2}\). Following the geometric mandates of Theorem \ref{t1}, a production hypersurface \(\mathcal{M}\) emerges as a stable, time-independent two-dimensional integral leaf of the involutive distribution \(\mathcal{D}_1 = \text{span}\{X_s, X_1\}\) if and only if it corresponds to the level set of a master invariant \(H_3 = c_1 H_1 + c_2 H_2\) that is simultaneously annihilated by both generators. Enforcing the scaling restriction 
\[X_s(H_3) = c_1\left(1 - \frac{b_3}{b_1}\right) + c_2\left(1 - \frac{b_3}{b_2}\right) = 0\] establishes the crucial structural tracking condition 
\begin{equation}
\label{tracking}
c_1 + c_2 = c_1\frac{b_3}{b_1} + c_2\frac{b_3}{b_2}.
\end{equation}
Expanding the zero-level set equation \(H_3 = \text{constant}\) in terms of extensive coordinates yields: $$(c_{1}+c_{2})\ln Y-\left(c_{1}\frac{b_{3}}{b_{1}}\right)\ln K-\left(c_{2}\frac{b_{3}}{b_{2}}\right)\ln L=\text{constant}.$$
Dividing through by the total weight \((c_1 + c_2)\) isolates the output variable as 
$$\ln Y = \alpha \ln K + \beta \ln L + \text{constant},$$
where the endogenously generated elasticity coefficients are uniquely determined by the dynamical growth rates as:
$$\alpha =\frac{c_{1}b_{3}}{b_{1}(c_{1}+c_{2})}\quad \text{and}\quad \beta =\frac{c_{2}b_{3}}{b_{2}(c_{1}+c_{2})}.$$
Summing these derived exponents and substituting the structural tracking condition (\ref{tracking}) into the numerator yields:
$$\alpha +\beta =\frac{1}{c_{1}+c_{2}}\left(c_{1}\frac{b_{3}}{b_{1}}+c_{2}\frac{b_{3}}{b_{2}}\right)=\frac{c_{1}+c_{2}}{c_{1}+c_{2}}\equiv 1.$$
Exponentiation of this isolated leaf equation recovers the classical Cobb–Douglas production function \(Y = AK^\alpha L^\beta\) endogenously, rigorously proving that constant returns to scale (CRTS) is not an empirical assumption but a direct geometric consequence of leaf alignment within the bi-Hamiltonian distribution \(\mathcal{D}_{1}\). Indeed, since $\beta = 1-\alpha$, we arrive at the CRTS Cobb-Douglas form (\ref{CD}).

Next, recall that in a 3D manifold equipped with a volume form \(\Omega \), any skew-symmetric contravariant bivector \(\pi \) is uniquely isomorphic to a differential 1-form 
\(\alpha \) via interior contraction:
\begin{equation}
\label{alpha}
\alpha =\iota _{\pi }\Omega. 
\end{equation}
The non-linear Schouten bracket condition for integrability (\ref{eq:poisson_integrability}) is mathematically equivalent to the Frobenius integrability condition for its dual 1-form (see Chapter 2 in Crainic {\em et al.} \cite{CFM2021}): 
$$\alpha \land d\alpha =0. $$
Let us compute this contraction explicitly using the coordinates $(x^1, x^2, x^3) = (K, L, Y)$ and the log-canonical volume form given in (\ref{logV}). Indeed, for any 3D bivector $$\pi = \pi^{12}\partial_K \wedge \partial_L + \pi^{13}\partial_K \wedge \partial_Y + \pi^{23}\partial_L \wedge \partial_Y,$$the contraction formula yields: 
\begin{equation}
\label{construction}
\iota _{\pi }\Omega _{\log }=\frac{1}{KLY}\left(\pi ^{23}\mbox{d}K-\pi ^{13}\mbox{d}L+\pi ^{12}\mbox{d}Y\right).
\end{equation}
Finally, by sequentially cross-contracting these Nambu invariants against our log-canonical volume form, we induce a compatible pair of  Poisson bivectors:
\begin{equation}
\label{pb}
{\pi }_{1}= \frac{b_1b_2}{b_3} \Omega _{\log }^{-1}(\mbox{d}H_{2}\land \cdot )\quad \text{and}\quad {\pi }_{2}=-\frac{b_1b_2}{b_3}\Omega _{\log }^{-1}(\mbox{d}H_{1}\land \cdot ).
\end{equation} 
Indeed, in the local coordinate basis $(\partial_K\wedge  \partial_L, \partial_K\wedge  \partial_Y, \partial_L\wedge  \partial_Y)$, these structures evaluate to the perfectly quadratic skew-symmetric matrices: 
\begin{equation}
\label{poisson}
\pi_1 = \begin{pmatrix}
0 & \frac{b_1 b_2}{b_3} KL & b_1 KY \\
-\frac{b_1 b_2}{b_3} KL & 0 & 0 \\
-b_1 KY & 0 & 0
\end{pmatrix}, \quad 
\pi_2 = \begin{pmatrix}
0 & b_2 KL & 0 \\
-b_2 KL & 0 & -\frac{b_1 b_2}{b_3} LY \\
0 & \frac{b_1 b_2}{b_3} LY & 0
\end{pmatrix}.
\end{equation}
Let us show now that the bivectors $\pi_1$ and $\pi_2$ are indeed Poisson bivectors. First, we construct the 1-form $\alpha_1$ with respect to bivector $\pi_1$ given by (\ref{poisson}), using (\ref{alpha}) and (\ref{construction}). We have $\pi^{12} = \frac{b_1b_2}{b_3}KL$, $\pi^{13} = b_1KY$, $\pi^{23} =0$. Substituting these components into the contraction formula: $$\alpha _{1}=\iota _{\pi _{1}}\Omega _{\log }=\frac{1}{KLY}\left\{(LY)\mbox{d}K-\left(\frac{b_{3}}{b_{2}}KY\right)\mbox{d}L+\left(\frac{b_{3}}{b_{2}}KL\right)\mbox{d}Y\right\}.$$
Canceling the coordinate variables, we arrive at the following exact 1-form: 
$$\alpha _{1}=\frac{\mbox{d}K}{K}-\frac{b_{3}}{b_{2}}\frac{\mbox{d}L}{L}+\frac{b_{3}}{b_{2}}\frac{\mbox{d}Y}{Y} = \mbox{d}\left( \ln K - \frac{b_3}{b_2}\ln L + \frac{b_3}{b_2}\ln Y \right),$$
which immediately implies integrability $\alpha_1\wedge \mbox{d}\alpha_1 = 0$, which in turn implies the Poisson integrability (\ref{eq:poisson_integrability}) for the bivector $\pi_1$. Therefore, we conclude that $\pi_1$ is indeed a Poisson bivector. Repeating the same argument for $\pi_2$ {\em mutatis mutandis}, we verify in a similar way that $\pi_2$ is also a Poisson bivector. To prove the compatibility (\ref{eq:poisson_compatibility}) for the Poisson bivectors $\pi_1$ and $\pi_2$ given by (\ref{poisson}), we employ the same argument rooted in  Poisson Geometry \cite{CFM2021}. 

To establish the algebraic compatibility of the Poisson structures \(\pi _{1}\) and \(\pi _{2}\) given by (\ref{poisson}), we must prove that any arbitrary linear combination (the Poisson pencil) \(\pi_\lambda = \pi_1 + \lambda \pi_2\) satisfies the Poisson integrability condition (\ref{eq:poisson_compatibility}) for all scalars \(\lambda \in \mathbb{R}\). Rather than deploying coordinate-dependent expansions of the Schouten bracket \cite{JS1940}, we leverage the native algebraic properties of the underlying Nambu–Poisson framework. Recall that the bivectors \(\pi _{1}\) and \(\pi _{2}\) are generated by cross-contracting the differentials of the uncoupled log-canonical invariants \(H_{1}\) and \(H_{1}\) as determined by (\ref{pb}). Constructing the linear pencil bivector explicitly yields:
$$\pi _{\lambda }=\pi _{1}+\lambda \pi _{2}=\frac{b_{1}b_{2}}{b_{3}}\Omega _{\log }^{-1}\left(\mbox{d}H_{2}\land \cdot \right)-\lambda \frac{b_{1}b_{2}}{b_{3}}\Omega _{\log }^{-1}\left(\mbox{d}H_{1}\land \cdot \right).$$
By the linearity of the vector-volume isomorphism \(\Omega _{\log }^{-1}\), we construct the linear pencil bivector explicitly. Rather than utilizing the unscaled differences of the background invariants, we map the combined matrix entries (\ref{poisson}) to a unified, parameter-dependent closed 1-form \(\alpha_\lambda = \iota_{\pi_\lambda}\Omega_{\log} = \alpha_1 + \lambda \alpha_2\). Integrating this combined 1-form across the continuous scalar field of \(\lambda \in \mathbb{R}\) yields: $$\alpha _{\lambda }=\mbox{d}\left\{\ln K-\frac{b_{3}}{b_{2}}\ln L+\frac{b_{3}}{b_{2}}\ln Y+\lambda \left(\frac{b_{3}}{b_{1}}\ln K-\ln L+\frac{b_{3}}{b_{1}}\ln Y\right)\right\}=\mbox{d}H_{\lambda },$$
where \(H_{\lambda }\) acts as the uniform master Hamiltonian function of the pencil. Because \(\alpha _{\lambda }\) is structurally recovered as an exact differential of a smooth scalar function (\(\alpha_\lambda = \mbox{d}H_\lambda\)), its exterior derivative vanishes identically:  \(\mbox{d}\alpha _{\lambda }=\mbox{d}^{2}H_{\lambda }\equiv 0.\)Under the rules of 3D Poisson-volume geometry, the closure of this dual 1-form trivially satisfies the vanishing condition across the state space, ensuring that \(\pi_\lambda = \pi_1 + \lambda\pi_2\) remains a valid, integrable Nambu–Poisson contraction for any choice of \(\lambda \). Next, expanding the Schouten self-bracket of the pencil via bilinearity yields:
$$[\pi _{\lambda },\pi _{\lambda }]=[\pi _{1}+\lambda \pi _{2},\;\pi _{1}+\lambda \pi _{2}]=[\pi _{1},\pi _{1}]+2\lambda [\pi _{1},\pi _{2}]+\lambda ^{2}[\pi _{2},\pi _{2}]\equiv 0.$$
Since \(\pi _{1}\) and \(\pi _{2}\) are independently verified to be valid Poisson structures, their individual Schouten self-brackets vanish identically (\([\pi_1, \pi_1] = 0\) and \([\pi_2, \pi_2] = 0\)). This leaves the single scalar parameter constraint:  \(2\lambda [\pi _{1},\pi _{2}]=0,\) \(\forall \lambda \in \mathbb{R}.\)  Because this relation must hold throughout the entire continuous parameter field of \(\lambda \), the cross-bracket tensor itself is forced to vanish: \([\pi _{1},\pi _{2}]=0.\) Therefore, the  Poisson bivectors \(\pi _{1}\) and \(\pi _{2}\) given by (\ref{poisson}) form a truly compatible bi-Hamiltonian pencil, anchoring the complete integrability of the unconstrained economic growth field \(X_{1}\), which is proven to be a bi-Hamiltonian vector field determined by two compatible Poisson bivectors. The bi-Hamiltonian structure in this case is rigid,  stable, and {\em unconstrained}. We also note that the functions \(H_{1}\) and \(H_{2}\) natively satisfy the interchanging duality of the bi-Hamiltonian architecture, where each function acts as the active Hamiltonian generating the growth field flow lines under its respective structure, while acting as the structural Casimir invariant spanning the kernel of the opposing structure. 

It must be mentioned that for the first time a bi-Hamiltonian architecture for the exponential growth model was previously introduced in \cite{SW2021}; however, that earlier formulation was constructed strictly with respect to the flat Euclidean volume form (\ref{eV}), which necessitated the macroeconomically restrictive parametric constraint \(b_1 + b_3 = b_2\) to maintain a volume-preserving flow. In contrast, the geometric framework deployed in this paper bypasses this limitation entirely by utilizing a scale-invariant log-canonical volume form (\ref{logV}), which accommodates independent, strictly positive growth rates (\(b_i > 0\)) without introducing artificial structural locks between the variables.

\begin{figure}[ht]
 \centering
  \includegraphics[width=10cm,height=8cm]{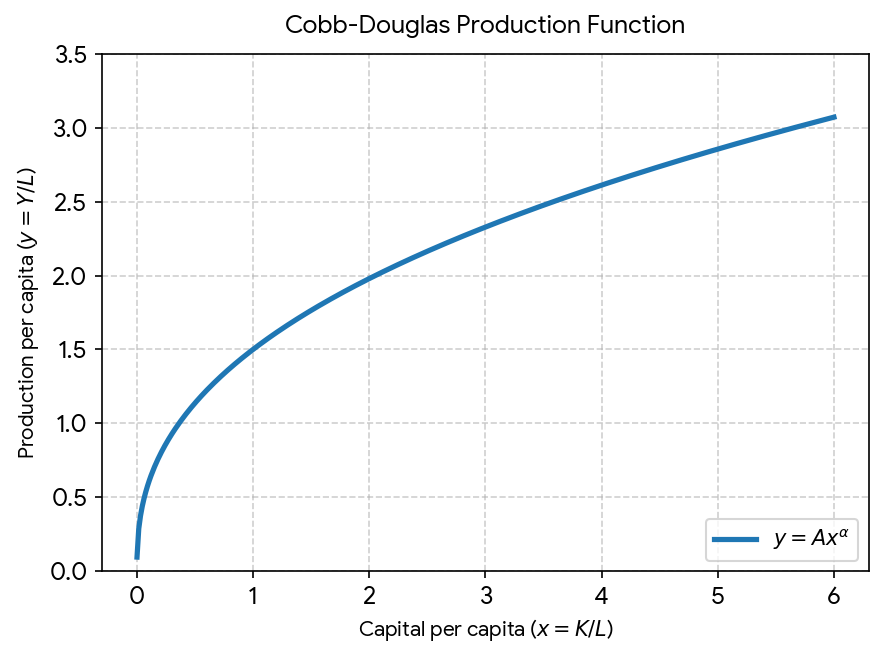}
   \caption{The one-input Cobb-Douglas production function.} 
   \label{figure1}
\end{figure}

The one-input factor Cobb-Douglas production function is illustrated by  Figure \ref{figure1}. 

\section{Case II: The S-Shaped (Sigmoidal) Regime}
\label{s42}

In this section, we investigate the next model from a bi-Hamiltonian perspective. This approach naturally generalizes the simultaneous holotheticity regime, which led to the derivation of the Cobb-Douglas production function in Section \ref{s41}. As shown in \cite{SV2026}, the underlying scale-invariant vector field can be employed to derive a certain $S$-shaped production function. This approach naturally generalizes the simultaneous holotheticity regime, which led to the derivation of the Cobb-Douglas production function (\ref{CD}) in Section~\ref{s41}.  This function, introduced {\em ad hoc}, has been widely studied in recent years within various models arising in economics, population dynamics, and ecology \cite{ACKL2013, ACKL2015, CEL2010, CEL2012, LLM2015, PWZ2026}. 

Consider now the integrable distribution introduced in \cite{SV2026} to model simultaneous holotheticity in this case: 
\begin{equation}
\label{D2}
\mathcal{D}_2 =\left\{X_s= K\frac{\partial}{\partial K} + L\frac{\partial}{\partial L} + Y\frac{\partial}{\partial Y}, \, X_2 = b_1K\frac{\partial}{\partial K} + b_2L\frac{\partial}{\partial L} +  b_3Y\left(1 - \frac{Y}{N_Y L}\right)\frac{\partial}{\partial Y}\right\}.
\end{equation}
We first note that the distribution (\ref{D2}) reduces to (\ref{D1}) as $N_Y \to \infty$. Therefore, we expect all of the structures associated with this distribution to reduce to the corresponding structures of the distribution (\ref{D1}) as $N_Y \to\infty$. 

In what follows, we will construct a bi-Hamiltonian structure and derive the corresponding production function for the vector field $X_2$ given by (\ref{D2}). First, following the algorithm  we employed in Section \ref{s41}, 
we compute 
\begin{equation}
\label{div2}
\text{div}_{\Omega _{\log }}(X_{2}) = -\frac{b_3 Y}{N_YL}.
\end{equation}
This calculation puts in evidence that the vector field \(X_{2}\) is not globally volume-preserving  on the  space \(M = \mathbb{R}_{+}^{3}\). Now, consider instead the following volume form compatible with (\ref{D2}): 
\begin{equation}
\label{sigV}
\Omega _{\mbox{\tiny sig}}=\frac{1}{K\cdot  L\cdot (N_Y- Y)}\mbox{d}K\land \mbox{d}L\land \mbox{d}Y.
\end{equation}
Re-evaluating the generalized  divergence under \(\Omega _{\mbox{\tiny sig}}\) yields:
\begin{equation}
\label{div3}
\text{div}_{\Omega _{\mbox{\tiny sig} }}(X_{2}) = 0.
\end{equation}
This confirms that the logistic vector field $X_2$ is globally unimodular with respect to the sigmoidal form $\Omega_{\mbox{\tiny sig}}$, and thus can be represented as a bi-Hamiltonian vector field in 3D within the framework of Nambu-Poisson mechanics. Indeed, let 
\begin{align*}
& H_1 = \frac{1}{N_Y}\ln\left(\frac{Y}{N_Y  - Y}\right) - \frac{b_3}{b_2N_Y}\ln L,  \\
& H_2 = b_2\ln K - b_1 \ln L. 
\end{align*}
Then a direct computation gives 
$$\iota_{X_2}\Omega_{\mbox{\tiny sig}} = -\mbox{d}H_1 \wedge \mbox{d}H_2.$$
Therefore, the vector field $X_2$ is the Nambu vector field associated with the pair of first integrals (Hamiltonians) $(H_1, H_2)$. Following the framework employed in Section \ref{s41}, we use the fact that a 3-form $\Omega$, one can define for each 1-form $\alpha$ the bivector $\pi$ such that
$$\iota_{\pi}\Omega = \alpha.$$
Define 
\begin{equation} 
\label{p12}
\pi_1:=\Omega^{-1}_{\mbox{\tiny sig}} (\mbox{d}H_2), \quad \pi_2:= \Omega^{-1}_{\mbox{\tiny sig}} (\mbox{d}H_1). 
\end{equation}
Hence
$$\iota_{\pi_1}\Omega_{\mbox{\tiny sig}} = \mbox{d}H_2, \quad \iota_{\pi_2}\Omega_{\mbox{\tiny sig}} = \mbox{d}H_1.$$
Since $\mbox{d}H_1$ and $\mbox{d}H_2$ are exact, the Frobenius integrability condition $\alpha \wedge \mbox{d}\alpha = 0$ trivially holds in both cases,  so both $\pi_1$ and $\pi_2$ satisfy the Poisson integrability condition (\ref{eq:poisson_integrability}) and, as such, are genuine Poisson bivectors. Next, recall that for any Poisson bivector $\pi$ and function $H$, the Hamiltonian vector field $X_H = [\pi, H]$ satisfies
$$\iota_{X_H}\Omega = -\mbox{d}H\wedge (\iota_{\pi}\Omega).$$
Therefore, with $\pi_1$  and the Hamiltonian $H_1$, we have
$$\iota_{[\pi_1, H_1]}\Omega_{\mbox{\tiny sig}} = - \mbox{d}H_1\wedge (i_{\pi_1}\Omega_{\mbox{\tiny sig}}) = -\mbox{d}H_1 \wedge \mbox{d}H_2 = i_{X_2}\Omega_{\mbox{\tiny sig}},$$
where $X_2 = [\pi_1, H_1]$, as expected. Similarly, it can be shown that $X_2 = [\pi_2, H_2]$. Therefore, the vector field $X_2$ enjoys the bi-Hamiltonian representation (\ref{eq:bihamiltonian_pencil}): 
$$X_2 = [\pi_1, H_1] = [\pi_2, H_2].$$
We also note that $H_1$ and $H_2$ are the Casimirs for the alternative Poisson bivectors: $[\pi_1, H_2] = [\pi_2, H_1] = {\bf 0}$. We can also verify, using a similar argument, that the Poisson bivectors $\pi_1$ and $\pi_2$ are, in fact,  compatible (\ref{eq:poisson_compatibility}). Indeed, the pencil $\pi_{\lambda} = \pi_1 + \lambda \pi_2$, $\lambda \in \mathbb{R}$ has dual 1-form
$$\iota_{\pi_{\lambda}} = \mbox{d}H_2 + \lambda \mbox{d}H_1.$$
Since both $\mbox{d}H_1$ and $\mbox{d}H_2$ are exact, the 1-form is closed for every $\lambda\in \mathbb{R}$, hence
$$(\mbox{d}H_2 + \mbox{d}H_1)\wedge \mbox{d}(\mbox{d}H_2 + \mbox{d}H_1) = 0.$$
Thus, $\pi_{\lambda}$ is a Poisson bivector for all $\lambda \in \mathbb{R}$, which implies the compatibility condition (\ref{eq:poisson_compatibility}). Furthermore, the Poisson bivectors $\pi_1$ and $\pi_2$ given by (\ref{p12}) enjoy the following explicit forms with respect to the basis ($\partial_K\wedge \partial_L$, $\partial_K\wedge \partial_Y$, $\partial_L \wedge \partial_Y$): 
$$
\pi_1 = - b_1KY(N_Y - Y)\partial_K \wedge \partial_Y, \quad \pi_2 = -  \frac{KL}{N_Y}\partial_K\wedge\partial_L - \frac{b_3KY(N_Y-Y)}{b_2N_Y}\partial_K\wedge \partial_Y. 
$$
We can adjust the signs by absorbing constants into the Hamiltonians; the essential point is that both are Poisson and compatible. Now, consider the following combination: 
$$N_YH_1 - \frac{p}{b_2}H_2 = C_1,$$
where $C_1 \in  \mathbb{R}$ is a constant. Substituting, we get
$$\left[\ln\left(\frac{Y}{N_Y - Y}\right) - \frac{b_3}{b_2}\ln L\right] -\frac{p}{b_3}[b_2\ln K - b_1\ln L].$$
Expanding and simplifying, we arrive at 
$$\ln\left(\frac{Y}{N_Y - Y}\right) - p\ln K + \left(\frac{pb_1 - b_3}{b_2}\right) \ln L  = C_1.$$
Setting $p  = \frac{b_3- b_2}{b_1 - b_2}$ and noticing that $\frac{pb_1 - b_3}{b_2} = \frac{b_3 - b_1}{b_1 - b_2} = p-1$, we obtain: 
$$\ln\left(\frac{Y}{N_Y - Y}\right) - p\ln K + (p-1)\ln L, $$
or, 
$$\ln\left(\frac{Y L^{p-1}}{(N_Y - Y)K^p}\right)  = C_1,$$
from which we get, solving for $Y$: 
\begin{equation}
\label{Sshaped}
Y = \frac{CN_YK^pL^{1-p}}{1 + CK^pL^{-p}}, \quad C_1 = \ln C. 
\end{equation}
Finally, by setting $\alpha_1 = CN >0$, $\alpha_2 = C>0$ in (\ref{Sshaped}), we finally arrive at the $S$-shaped production function form studied in \cite{ACKL2013, ACKL2015, CEL2010, CEL2012, LLM2015, PWZ2026}: 
\begin{equation}
\label{Sshaped1}
Y = F(K, L) = \frac{\alpha_1 K^pL^{1-p}}{1 + \alpha_2K^pL^{-p}}.
\end{equation}

Alternatively, we can derive the $S$-shaped production form (\ref{Sshaped1}) by projecting the distribution \(\mathcal{D}_{2}\) given by (\ref{D1}) onto the two-dimensional projective per-capita manifold \(M_{\text{proj}} = \{(x,y) \in \mathbb{R}^2_+\}\) via the projective coordinates \(x = K/L\) and \(y = Y/L\). Applying the quotient rule to the components of the vector field $X_2$ given by (\ref{D2}), the autonomous projected growth field 
${X}_{2}$ evaluates to (see \cite{SV2026} for more details): 
$$\widetilde{X}_{2}=(b_{1}-b_{2})x\frac{\partial }{\partial x}+(b_3-b_2)y\left[1 -\frac{y}{N_{Y}}\right]\frac{\partial }{\partial y},$$ 
where $N_y = \left(1 - \frac{b_2}{b_3}\right)N_Y$.  We begin by constructing a Hamiltonian structure for the vector field $\tilde{X}_2$. To accomplish this task, we utilize the Hamiltonian structure found in \cite{SW2019} for a more general ``logistic" 3D model.  This yields the following  Poisson bivector:
\begin{equation}
\tilde{\pi} = -(b_1 - b_2)xy\left(1 - \frac{y}{N_y}\right) \frac{\partial}{\partial x} \wedge \frac{\partial}{\partial y}.
\label{eq:airtight_sigmoidal_pencil}
\end{equation}
Since the system is defined on a 2-dimensional manifold, any tri-vector field must identically vanish because \(\Lambda^3(T\mathbb{R}^2) = \{0\}\). Consequently, the Schouten bracket \([\tilde{\pi}, \tilde{\pi}]\) is automatically zero, meaning the skew-symmetric bivector (\ref{eq:airtight_sigmoidal_pencil}) trivially satisfies the Jacobi identity and is a valid Poisson bivector. The corresponding Hamiltonian is found to be
\begin{equation}
h(x, y) = \ln \left(\frac{y}{N_y - y}\right) -  \left(\frac{b_3 - b_2}{b_1 - b_2}\right) \ln x, \label{eq:airtight_h1} 
\end{equation}
from which we find $\widetilde{X}_2 = [\tilde{\pi}, h]$. 
Because the background area profile is treated via a single structural layout rather than forcing a multi-Hamiltonian coordinate lock (which is structurally overdetermined on a 2D surface), the level-leaf equation \(h(x,y) = C\) resolves without any transcendental roadblocks:
$$\ln \left(\frac{y}{N_{y}-y}\right)-p\ln x=C_1,$$ 
where $p=\frac{b_{3}-b_{2}}{b_{1}-b_{2}}$ and $C_1 \in \mathbb{R}$ is a constant. Solving for $y$, we arrive at the one-input version of the $S$-shaped production function (\ref{Sshaped1}) given by 
$$
y = f(x) = \frac{N_yCx^p}{1+Cx^p}, \quad C_1 = \ln C.
$$
Renaming the parameters in the above as $\alpha_1 = N_yC>0$ and $\alpha_2  = C>0$, we arrive at the well-known form of the one-input version of the $S$-shaped (sigmoidal) production function (\ref{Sshaped1}): 
\begin{equation}
\label{Sshaped2}
y = f(x) = \frac{\alpha_1x^p}{1+\alpha_2x^p}. 
\end{equation}

Indeed, pulling this result back by substituting the coordinate fractions \(x = K/L\) and \(y = Y/L\) and multiplying through by labor \(L\) recovers the exact extensive $S$-shaped production function (\ref{Sshaped1}). Thus, the sigmoidal production manifold is structurally vindicated as an exact linear leaf of a 2D Hamiltonian architecture.

It is noteworthy that the functional form given in equation (\ref{Sshaped2}) is structurally identical to the well-known Holling Type III functional response from ecology, which describes the rate of prey consumption by a predator as a function of prey density. The conceptual framework for functional responses was originally developed by a Canadian ecologist C.~S. ``Buzz" Holling, who classified them into three types --- I, II, and III --- based on the shape of the relationship between feeding rate and prey abundance \cite{Holling1, Holling2, Holling3}. While Holling originally described the Type III response qualitatively as a sigmoid curve, the specific mathematical formulation given by (\ref{Sshaped2}) was later proposed by L.~A.~Real \cite{Real1977}, who derived the equation through an analogy with enzyme kinetics. Real's general functional response model provided a unified framework that could account for both Type II and Type III responses --- see also \cite{Denny2014}. 

\begin{figure}[ht]
 \centering
  \includegraphics[width=10cm,height=8cm]{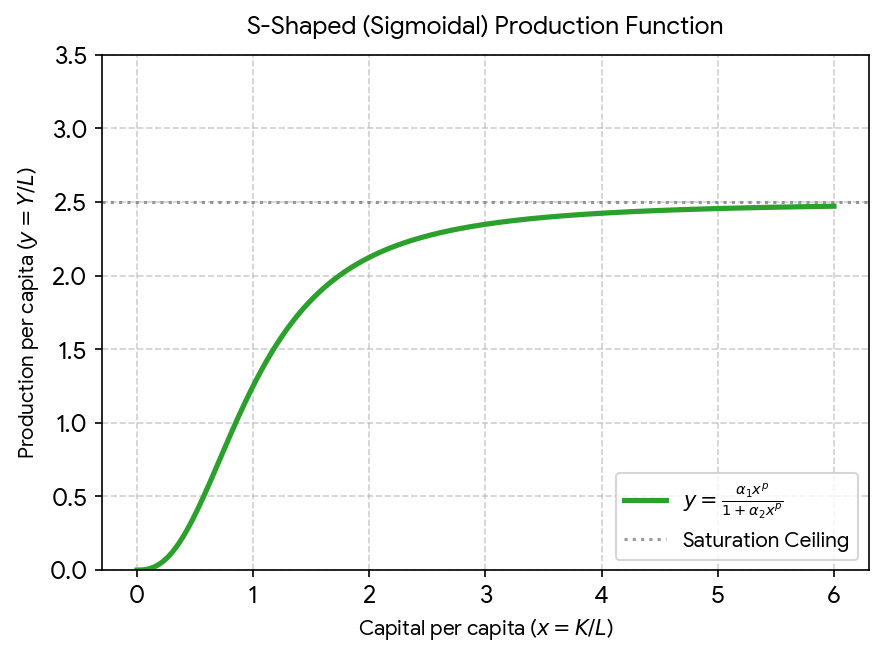}
   \caption{The one-input $S$-shaped production function.} 
   \label{figure2}
\end{figure}

The one-input factor $S$-shaped  production function is illustrated by  Figure \ref{figure2}. 

\section{The “Overshoot-and-Collapse” Regime and the Breakdown of the Smooth Bi-Hamiltonian Structure}
\label{s43}

In this section, we investigate from the bi-Hamiltonian perspective the final and most complex model within our hierarchical geometric architecture: the ``overshoot-and-collapse" economic regime introduced in \cite{SV2026}. This regime represents a realistic macroeconomic scenario where both capital accumulation and production output are constrained by non-smooth carrying capacities. The simultaneous holotheticity of this regime is governed by the integrable distribution \(\mathcal{D}_3 = \{X_s, X_3\}\) given by 
\begin{equation}
\label{D3}
\mathcal{D}_3 = \left\{ 
\begin{aligned}
X_s &= K \frac{\partial}{\partial K} + L \frac{\partial}{\partial L} + Y \frac{\partial}{\partial Y}, \\[1ex]
X_3 &= b_1K \left( 1 - \frac{K}{N_K L} \right) \frac{\partial}{\partial K} + b_2L \frac{\partial}{\partial L} + b_3Y \left( 1 - \frac{Y}{N_Y L} \right) \frac{\partial}{\partial Y}.
\end{aligned}
\right\}
\end{equation}
We first note that the distribution $\mathcal{D}_3$ given by (\ref{D3}) reduces to $\mathcal{D}_2$ (\ref{D2}) as $N_K \to\infty$ and to --- $\mathcal{D}_1$ (\ref{D1}) as both $N_K\to\infty$ and $N_Y\to\infty$. Therefore, we have a stacked hierarchy of integrable distributions describing three different economic regimes that are inherently related. We also note that the vector field $X_3$ is of the type determined by the formula (\ref{genX}), as expected.

Following the precise algorithmic sequence established in Sections \ref{s41} and \ref{s42}, we first evaluate the structural volume properties of the evolutionary vector field \(X_{3}\). Computing the generalized divergence under the scale-invariant log-canonical volume form \(\Omega _{\log }\) given in equation (\ref{logV}) yields: $$\text{div}_{\Omega _{\log }}(X_{3})=-\frac{b_{1}K}{N_{K}L}-\frac{b_{3}Y}{N_{Y}L}.$$
This non-vanishing result explicitly shows that the vector field \(X_{3}\) is not globally volume-preserving or unimodular on the open orthant \(\mathbb{R}_{+}^{3}\) under \(\Omega _{\log }\). To restore the volume-preserving Nambu-Poisson architecture, we apply a dual conformal gauge transformation that simultaneously accounts for the capacity constraints bounding both capital and production. This induces the modified log-logistic volume form specific to the collapse regime:
\begin{equation}
\label{volC}
\Omega _{\text{cusp}}=\frac{1}{L\cdot|N_{K}L-K|\cdot|N_{Y}L-Y|}\mbox{d}K\land \mbox{d}L\land \mbox{d}Y.
\end{equation}

Re-evaluating the generalized divergence under this sigmoidal capacity volume form yields $$\text{div}_{\Omega_{\text{cusp}}}(X_3) = 0,$$
which confirms that under this volume form, the system is perfectly unimodular. Instead of recovering the system phase velocities from transcendental or parametrically locked logarithmic expressions, we find the requisite first invariants by solving the characteristic system of $X_3$:
\begin{equation}
\frac{\mbox{d}K}{b_1 K \left(1 - \frac{K}{N_K L}\right)} = \frac{\mbox{d}L}{b_2 L} = \frac{\mbox{d}Y}{b_3 Y \left(1 - \frac{Y}{N_Y L}\right)}
\end{equation}
Inverting the linearizing coordinates and evaluating the quadratures eliminates the explicit time wrapper $t$, yielding two functionally independent, streamlined master invariants $H_1$ and $H_2$:
\begin{align}
H_1(K, L, Y) &= L^{\frac{b_1-b_2}{b_2}} \left( \frac{L}{K} - \frac{b_1}{N_K(b_1-b_2)} \right), \label{eq:I1_inv} \\
H_2(K, L, Y) &= L^{\frac{b_3-b_2}{b_2}} \left( \frac{L}{Y} - \frac{b_3}{N_Y (b_3-b_2)} \right). \label{eq:I2_inv}
\end{align}
 Direct computation verifies that their directional derivatives vanish identically ($X_3(H_1) = 0$, $X_3(H_2) = 0$) across the open quadrant for arbitrary growth rates ($b_1 \neq b_2 \neq b_3$), satisfying the definition of constants of motion without forcing parameter constraints.

Cross-contracting the invariants (\ref{eq:I1_inv})-(\ref{eq:I2_inv}) against the background  volume form (\ref{volC}) induces a compatible pair of contravariant Poisson bivectors $\pi_1 := \Omega_{\text{cusp}}^{-1}(\mbox{d}H_2 \wedge \cdot)$ and $\pi_2 := -\Omega_{\text{cusp}}^{-1}(\mbox{d}H_1 \wedge \cdot)$. Expressed in the local coordinate basis $(\partial_K \wedge \partial_L, \, \partial_K \wedge \partial_Y, \, \partial_L \wedge \partial_Y)$, these structures simplify to:
\begin{equation}
\label{p11}
\pi_1 = 
\begin{pmatrix}
0 & -b_1 K L D_K & b_1 K L D_K\left(1-\dfrac{K}{N_K L}\right) \\[6pt]
b_1 K L D_K & 0 & b_3 Y L D_K\left(1-\dfrac{Y}{N_Y L}\right) \\[6pt]
-b_1 K L D_K\left(1-\dfrac{K}{N_K L}\right) & -b_3 Y L D_K\left(1-\dfrac{Y}{N_Y L}\right) & 0
\end{pmatrix},
\end{equation}
\begin{equation}
\label{p22}
\pi_2 = 
\begin{pmatrix}
0 & b_2 K L D_Y & b_1 K L D_Y\left(1-\dfrac{K}{N_K L}\right) \\[6pt]
-b_2 K L D_Y & 0 & b_2 L^2 D_Y \\[6pt]
-b_1 K L D_Y\left(1-\dfrac{K}{N_K L}\right) & -b_2 L^2 D_Y & 0
\end{pmatrix},
\end{equation}
where $D_K = |D_KL - K|$, $D_Y: = |D_YL - Y|$. These bivectors determine $X_3$ via $H_1$ and $H_2$ respectively,  satisfying $X_3 = [\pi_1, H_1] = [\pi_2, H_2]$,  alongside the Casimir constraints $[\pi_1, H_2] = [\pi_2, H_1]=\mathbf{0}$. Because they map directly onto closed 1-forms, their linear pencil remains integrable, satisfying the Schouten bivector integrability conditions (\ref{eq:poisson_integrability}) and (\ref{eq:poisson_compatibility}): $[\pi_1, \pi_1] = [\pi_2, \pi_2] = [\pi_1, \pi_2] = 0$ identically for all growth configurations \cite{CFM2021}. Therefore, we conclude that $\pi_1$ and $\pi_2$ given by (\ref{p11}) and (\ref{p22}) determine a bi-Hamiltonian structure for the vector field $X_3$ (\ref{D3}). 

Following the leaf alignment theorem, a stable macroeconomic production function hypersurface emerges as a mutual level set where a specific weighted linear combination balances out the remaining labor scaling components. Setting $p = \frac{b_3 - b_2}{b_1 - b_2}$, we combine the invariants via a power-transformed mapping $H_2 - \mathcal{C} H_1^p = 0$, where $\mathcal{C}$ is a constant.  Isolating the output state variable $Y$  yields the  ``overshoot-and-collapse" production function:
\begin{equation}
Y = F(K, L) = \frac{N_Y \mathcal{C} K^p L}{|N_K L - K|^p + \mathcal{C} K^p}. \label{eq:cuspext}
\end{equation}
By projecting the 3D phase flow onto the 2D projective per-capita orbit space via the canonical invariants $x = K/L$ and $y = Y/L$, the system collapses to the intensive form derived for the first time in \cite{SW2020} within a different context:
\begin{equation}
y = f(x) = \frac{N_y x^p}{x^p + \Omega |N_x - x|^p},  \label{eq:cuspint}
\end{equation}
where $N_x = \frac{b_1 - b_2}{b_1} N_K$ and $N_y = \frac{b_3 - b_2}{b_3} N_Y$. This structural derivation confirms that the ``overshoot-and-collapse" production profile is a native, geometric feature arising directly from the underlying Nambu matrix projections of the growth framework.

Alternatively, we can derive the production function (\ref{eq:cuspint}) by projecting the 3D phase flow onto the 2D projective per-capita manifold \(\mathcal{M}_{\text{proj}} = \{(x, y) \in \mathbb{R}^2_+\}\) via \(x = K/L\) and \(y = Y/L\). Applying the quotient rule reveals the autonomous projected growth field \(\widetilde{X}_{3}\) studied for the first time in \cite{SW2020}: 
\begin{equation}
\widetilde{X}_{3}=(b_{1}-b_{2})x\left(1-\frac{x}{N_{x}}\right)\frac{\partial }{\partial x}+(b_{3}-b_{2})y\left(1-\frac{y}{N_{y}}\right)\frac{\partial }{\partial y},
\end{equation}
where the intensive scaled capacities are given by \(N_x = \frac{b_1 - b_2}{b_1} N_K\) and \(N_y = \frac{b_3 - b_2}{b_3} N_Y\). This field maps onto a 2D Hamiltonian architecture \(\widetilde{X}_3 = [\tilde{\pi}, h]\) equipped with the  Poisson bivector $\tilde{\pi}$ given by
$$
\tilde{\pi} = -(b_1 - b_2) xy \left(1 - \frac{x}{N_x}\right)\left(1 - \frac{y}{N_y}\right)\partial_x \wedge \partial_y.
$$
Integrating the resulting coordinate gradients isolates the transcendental per-capita leaf equation: 
$$
h(x,y)=\ln \frac{|N_{y}-y|}{y}+p\ln \frac{x}{|N_{x}-x|}=C_{1},
$$
where \(p = \frac{b_3 - b_2}{b_1 - b_2}\). Setting \(C_1 = \ln C\), solving for \(y = f(x)\), we arrive at the following functional form, derived for the first time in \cite{SW2020} by a different method and in a different context: 
\begin{equation}
\label{lgen}
y = f(x) = \frac{N_y x^p}{x^p + \Omega |N_x - x|^p}.
\end{equation}
Next, pulling back to extensive parameters, we endogenously recover the capacity-limited production function for Case III introduced in \cite{SV2026} and given by (\ref{eq:cuspext}).

Importantly, the smooth bi-Hamiltonian complete integrability of Case III encounters a structural boundary crisis when the aggregate economic trajectories reach the resource carrying capacity ceilings.
Consider the switching manifold defined by the production capacity ceiling:
\[
\Sigma_Y = \{(K,L,Y)\in \mathbb{R}_+^3 \mid h_\Sigma(K,L,Y) = 0\},\qquad
h_\Sigma(K,L,Y) := N_Y L - Y.
\]

The state space is partitioned into $\mathcal{M}^- = \{Y < N_Y L\}$ and $\mathcal{M}^+ = \{Y > N_Y L\}$. The piecewise vector fields are defined as:
\begin{equation}
\label{X-}
X^- = b_1K\left(1 - \frac{K}{N_K L}\right)\frac{\partial}{\partial K}
      + b_2L\,\frac{\partial}{\partial L}
      + b_3Y\left(1 - \frac{Y}{N_Y L}\right)\frac{\partial}{\partial Y},
\end{equation}
\begin{equation}
\label{X+}
X^+ = b_1K\left(1 - \frac{K}{N_K L}\right)\frac{\partial}{\partial K}
      + b_2L\,\frac{\partial}{\partial L} 
      + \tilde{c}\, L\,\frac{\partial}{\partial Y},
\end{equation} 
where $\tilde{c} > 0$ is a dimensionless collapse rate parameter --- see below for an economic interpretation of this arrangement ----- note $X^+$ is not a projection of $X_3$ onto ${\cal M}^+$. This form preserves the scaling symmetry of the model, satisfying $[X_s, X^+] = 0$ with $X_s = K\partial_K + L\partial_L + Y\partial_Y$.

The normal vector to the switching manifold is $\nabla h = (0, N_Y, -1)$. Projecting both vector fields along this normal yields the directional derivatives on $\Sigma_Y$:
\[
X^-(h)\big|_{\Sigma_Y} = b_2 N_Y L > 0,
\]
\[
X^+(h)\big|_{\Sigma_Y} = (b_2 N_Y - \tilde{c}) L.
\]

The transversality condition for a Filippov sliding regime requires $X^-(h) > 0$ and $X^+(h) < 0$, which is satisfied if and only if
\begin{equation}
\label{cond}
\tilde{c} > b_2 N_Y >0.
\end{equation}

The Filippov convex combination is $X_{\text{fil}} = (1-\alpha)X^- + \alpha X^+$, where $\alpha \in (0,1)$ is the convex parameter. Requiring tangency along the switching manifold, $X_{\text{fil}}(h) = 0$, yields:
\[
\alpha = \frac{X^-(h)}{X^-(h) - X^+(h)}
= \frac{b_2 N_Y L}{(b_2 N_Y L) - (b_2 N_Y - \tilde{c}) L}
= \frac{b_2 N_Y}{\tilde{c}}.
\]

Since the transversality condition guarantees (\ref{cond}), it follows immediately that
\[
{0 < \alpha < 1.}
\]

Thus, the convex parameter is explicitly bounded in the interval $(0,1)$, ensuring a genuine Filippov sliding regime. Substituting $\alpha$ into the convex combination gives the sliding vector field on $\Sigma_Y$:
\begin{equation}
\label{fil}
X_{\text{fil}} = b_1K\left(1 - \frac{K}{N_K L}\right)\frac{\partial}{\partial K}
      + b_2L\,\frac{\partial}{\partial L}
      + b_2 N_Y L\,\frac{\partial}{\partial Y}.
\end{equation}

Since on $\Sigma_Y$ we have $Y = N_Y L$, this can equivalently be written as
\[
X_{\text{fil}} = b_1K\left(1 - \frac{K}{N_K L}\right)\frac{\partial}{\partial K}
      + b_2L\,\frac{\partial}{\partial L}
      + b_2 Y\,\frac{\partial}{\partial Y}.
\]

This vector field preserves the constraint manifold, as
\[
X_{\text{fil}}(h_\Sigma) = X_{\text{fil}}(N_Y L - Y) = N_Y(b_2 L) - b_2 N_Y L = 0,
\]
confirming that the sliding dynamics remain confined to the production capacity ceiling. 

We pause to offer an economic interpretation of the linear collapse given by
\begin{equation}
\label{linpf}
 Y = N_Y L   
\end{equation}
 that occurs on the switching manifold \(\Sigma_Y\). When the production function collapses to the linear form (\ref{linpf}), the economic implications are profound: output depends only on labor input in simple proportion, while capital \(K\) ceases to play any independent role in production. This represents the degeneration of a complex modern economy --- capital-intensive, monetized, and technologically sophisticated --- into a primitive ``natural economy" reliant solely on direct human labor for survival.

From a theoretical standpoint, the regression to the linear form \eqref{linpf} represents a macro-structural phase transition within a capacity-constrained system. In a complex, technologically sophisticated economy, stable monetary frameworks and institutional mechanisms serve as the essential order parameters that coordinate the division of labor and enable capital accumulation. When extreme systemic shocks or hyperinflationary crises destroy these coordination mechanisms, the order parameter collapses, forcing a rapid de-monetization and ``naturalization" of the economic structure. In this regime, the productive contribution of accumulated capital ceases to function independently, and the system undergoes a violent devolution into a subsistence-oriented mode of production where aggregate output scales purely linearly as a function of direct human labor input \cite{Clarke1999}.

The exterior vector field $X^+$  given by (\ref{X+}) represents a regulatory intervention or structural break that scales with the size of the economy (labor), ensuring that the policy response remains proportional to the scale of the system. The condition $\tilde{c} > b_2 N_Y$ guarantees that the intervention is sufficiently strong to overcome the natural growth of the capacity ceiling, independent of the state variables.

Crucially, the sliding vector field $X_{\text{fil}}$ given by (\ref{fil}) constructed above inherits a rigorous bi-Hamiltonian structure from the original compatible Poisson pencil. Since the switching function $h_\Sigma$ is a Casimir invariant of $\pi_2$, the Dirac bracket reduction $\pi_{\mathrm{Dirac}}$ is well-defined on the constraint manifold $\Sigma_Y$. A direct computation confirms that $X_{\text{fil}}$ is generated by the reduced Hamiltonian $H_1|_{\Sigma_Y}$ with respect to this Dirac bracket:
\[
X_{\text{fil}}= [\pi_{\mathrm{Dirac}}, H_1|_{\Sigma_Y}].
\]

Moreover, because the compatibility condition $[\pi_1,\pi_2] = 0$ is preserved under Dirac reduction, the reduced Poisson pencil $\pi_{\lambda}^{\mathrm{Dirac}} = \pi_1^{\mathrm{Dirac}} + \lambda \pi_2^{\mathrm{Dirac}}$ remains integrable for all $\lambda\in\mathbb{R}$. Hence, the sliding flow admits the bi-Hamiltonian representation
\[
X_{\text{fil}} = [\pi_1^{\mathrm{Dirac}}, H_1|_{\Sigma_Y}] = [\pi_2^{\mathrm{Dirac}}, H_2|_{\Sigma_Y}],
\]
on the codimension-one switching manifold --- see Chapter 7 in Crainic {\em et al.}  \cite{CFM2021} for more details. This demonstrates that the Filippov convex reconstruction does not destroy the geometric integrability of the system; rather, it projects the 3D bi-Hamiltonian hierarchy onto a lower-dimensional constraint leaf, preserving the algebraic duality of the Poisson structures. The Dirac reduction therefore provides a rigorous geometric mechanism for the collapse of the 3D dynamics onto the 2D sliding surface, formalizing the structural lock scenario without introducing mathematical singularities.

\begin{figure}[ht]
 \centering
  \includegraphics[width=10cm,height=8cm]{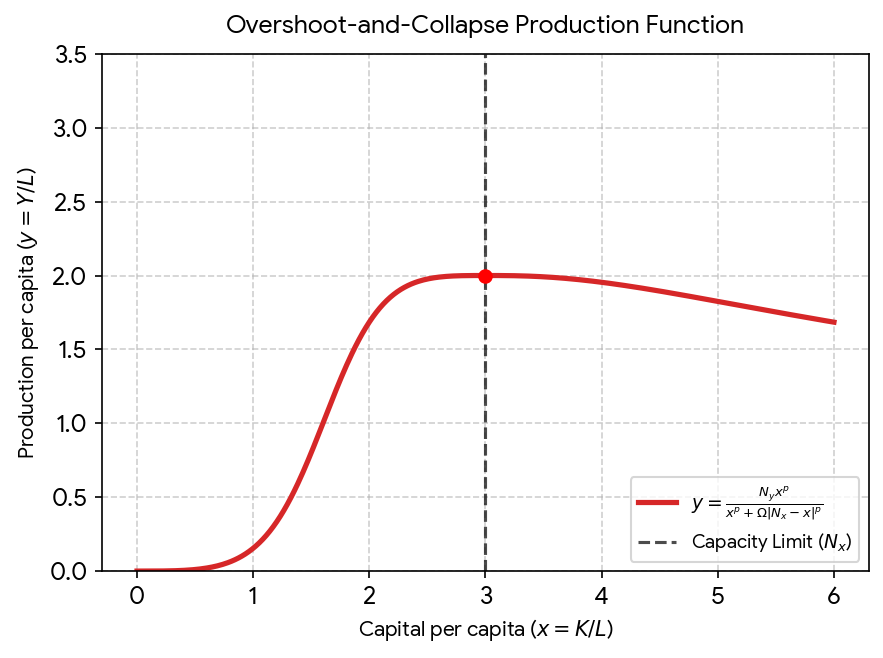}
   \caption{The one-input ``overshoot-and-collapse" production function.} 
   \label{figure3}
\end{figure}

The one-input factor ``overshoot-and-collapse" production function is illustrated by  Figure \ref{figure3}. 
 
\section{Conclusions} 
\label{s5}

In this paper, we have advanced the ongoing data-driven dynamical systems program in macroeconomic growth theory by elevating the analysis of production functions from empirical correlations to intrinsic geometric invariants. By extending the framework of continuous Lie groups, simultaneous holotheticity, and Nambu-Poisson mechanics into non-smooth configuration spaces, we have successfully mapped out the structural boundaries of scale-invariant phase flows across three distinct economic regimes: the unconstrained Cobb--Douglas framework, the $S$-shaped (sigmoidal) resource-limited ecosystem response, and the capacity-bounded overshoot-and-collapse regime. Table \ref{tab:economic_regimes}  describes their respective one-input factor functional forms. See also Figures \ref{figure1}, \ref{figure2}, and \ref{figure3}.

\begin{table}[htbp]
    \centering
    \caption{Structural Comparison of the Three Bi-Hamiltonian Economic Regimes}
    \label{tab:economic_regimes}
    \small
    \begin{tabular}{p{3.2cm} p{4.2cm} p{4.5cm}}
        \toprule
        \textbf{Regime Model} & \textbf{Mathematical Function} & \textbf{Macroeconomic Meaning} \\ 
        \midrule
        \textbf{Cobb-Douglas Regime} & $y = A x^{\alpha}$ & \textbf{Unconstrained growth} within the framework of the Inada conditions \cite{I1963}.  \\
        \addlinespace
        \textbf{S-Shaped Regime} & $y = \dfrac{\alpha_1 x^p}{1 + \alpha_2 x^p}$ & \textbf{Resource-limited response} where technology/ecology caps output asymptotically. \\
        \addlinespace
        \textbf{``Overshoot-and-Collapse" Regime} & $y = \dfrac{N_y x^p}{x^p + \Omega |N_x - x|^p}$ & \textbf{Capacity-bounded collapse} modeling systemic crashes and structural locks. \\
        \bottomrule
    \end{tabular}
\end{table}

Specifically, we have presented a unifying theory that combines, from a bi-Hamiltonian perspective, three fundamental growth regimes in economics. The first is the classical Cobb–Douglas regime, which has been extensively studied in the literature; it was initially introduced {\em ad hoc} \cite{H1997}, subsequently validated by fitting it to real empirical data \cite{CD1928, PD1976}, and eventually derived as a time-independent invariant by Sato \cite{RS1981}. The second is the $S$-shaped regime, which has traditionally been introduced {\em ad hoc} and studied extensively across various disciplines \cite{ACKL2013, ACKL2015, CEL2010, CEL2012, Holling1, Holling2, Holling3, LLM2015, PWZ2026, Real1977}; here, we have established it as a time-independent invariant of a dynamical system that directly generalizes the framework generating the Cobb–Douglas function. Finally, the ``overshoot-and-collapse" regime introduces a production function formalized as a time-independent invariant of a dynamical system that naturally generalizes the first two regimes. Crucially, all three production functions exist within a stacked structural hierarchy of sequential generalizations, where the ``overshoot-and-collapse" function generalizes the $S$-shaped profile, which in turn generalizes the foundational Cobb–Douglas model.

Our core geometric contribution lies in the resolution of the smooth transversality breakdown that occurs when aggregate economic trajectories reach resource carrying capacity boundaries. We have demonstrated that capacity limitations trigger a catastrophic rank-collapse of the compatible contravariant Poisson pencil $(\pi_1, \pi_2)$. By deploying Filippov's convex multiplier method on the codimension-one switching manifold $\Sigma$, we proved that the sliding attractor framework remains well-defined and non-singular, with the convex parameter $\alpha$ strictly bounded within the stable interval $(0, 1)$ under the parametric constraint $\tilde{c} > b_2 N_Y$.

A fundamental limitation of the classical bi-Hamiltonian theory, established by Fernandes in his seminal 1994 characterization \cite{RLF1994}, is that a bi-Hamiltonian structure defined by two globally defined non-degenerate Poisson bivectors is an exceptionally rigid construction. Fernandes proved that, under natural hypotheses, a compatible second non-degenerate Poisson bivector exists in a neighborhood of the invariant torus if and only if the graph of the Hamiltonian function $H(I)$ is a hypersurface of translation with respect to the affine structure generated by the action coordinates. Subsequent research by Boualem and Brouzet \cite{BB2021} further demonstrated that the set of functions that are additively separable after an affine transformation is topologically meager, implying that the vast majority of completely integrable Hamiltonian systems \emph{cannot} admit a standard bi-Hamiltonian representation with two globally non-degenerate Poisson bivectors. This rigidity poses a significant obstacle to applying the bi-Hamiltonian formalism to generic integrable systems, particularly those arising in economic growth theory where the underlying phase flows are often governed by degenerate or singular Poisson bivectors.

The Filippov sliding construction developed in Section \ref{s43} of this paper provides a systematic resolution to this fundamental limitation. Rather than requiring two globally non-degenerate Poisson bivectors on the full three-dimensional phase space, we construct a bi-Hamiltonian hierarchy that is defined \emph{piecewise} on the regular leaves of the phase space, with the transition between regimes mediated by Filippov's convex reconstruction \cite{F1988} on the switching manifold $\Sigma_Y$. The Dirac bracket reduction $\pi_{\mathrm{Dirac}}$ yields a well-defined Poisson bivector on the codimension-one constraint manifold, and the sliding vector field $X_{\mathrm{fil}}$ admits the bi-Hamiltonian representation (\ref{eq:bihamiltonian_pencil}),
where the reduced Poisson pencil $\pi_{\lambda}^{\mathrm{Dirac}} = \pi_1^{\mathrm{Dirac}} + \lambda \pi_2^{\mathrm{Dirac}}$ remains Poisson compatible (\ref{eq:poisson_compatibility})  for all $\lambda\in\mathbb{R}$.

Crucially, the Poisson bivectors $\pi_1$ and $\pi_2$ in the original 3D space are \emph{degenerate}, with Casimir invariants corresponding precisely to the first integrals $H_1$ and $H_2$ of the phase flow. This degeneracy, far from being a defect, is the essential geometric feature that allows the theory to bypass Fernandes' rigidity theorem: the compatible Poisson pencil need not be globally non-degenerate, and the bi-Hamiltonian structure is defined only on the symplectic leaves of the foliation, which in our economic setting correspond to the invariant production hypersurfaces. When the system approaches a resource capacity ceiling, the non-smooth transition is handled by the Filippov reconstruction, which effectively ``projects'' the degenerate bi-Hamiltonian hierarchy onto a lower-dimensional constraint leaf where the reduced Poisson bivectors become non-degenerate (up to a conformal factor) and the sliding flow is completely integrable in the Liouville--Arnold sense.

Thus, the Filippov sliding construction extends the bi-Hamiltonian theory to a broad class of degenerate Hamiltonian systems with singularities, including those arising from capacity-constrained economic growth models. This demonstrates that the topological meagreness obstruction identified by Fernandes and by Boualem $\&$ Brouzet can be overcome by allowing the Poisson bivectors to be degenerate and defined piecewise, with the non-smooth transitions handled by the convex reconstruction of Filippov. The present work therefore establishes a rigorous geometric framework for analyzing integrable systems that fall outside the classical bi-Hamiltonian characterization, providing a principled foundation for studying structural breaks, complexity ceilings, and systemic crises in economic growth theory.

From a macroeconomic perspective, this projection formalizes a rigorous geometric foundation for economic complexity ceilings and structural locks. When an economy hits its definitive capacity ceilings, the collapse of the production function onto a lower-dimensional leaf dictates a macro-structural phase transition: capital accumulation becomes entirely bound to labor dynamics, causing a sophisticated monetary economy to de-monetize and devolution into a linear mode of production scaling purely on direct human labor input ($Y = N_Y L$). 

Future research will focus on leveraging these explicit, quadratic Poisson bivector to design structure-preserving numerical integrators --- such as specialized average vector field (AVF) or Kahan splitting algorithms --- to simulate non-smooth economic phase flows over extended horizons without artificial dissipation. Additionally, extending this multi-Hamiltonian pencil layout to model multi-sector economic competitive environments via fractional Poisson bivector with history-dependent memory invariants represents a highly promising avenue of geometric investigation \cite{S2026}.

\subsection*{Acknowledgements}

The second author (RGS) wishes to express his deepest and most enduring gratitude to his late PhD supervisor, Professor Oleg Bogoyavlenskij (1948–2024). It was through his profound geometric intuition, uncompromising mathematical rigor, and encyclopedic knowledge that the author was first introduced to the beautiful and intricate theory of bi-Hamiltonian systems. 

A foundational portion of the author's mathematical perspective was shaped as a regular participant in Professor Bogoyavlenskij's research seminar during the 1990s at Queen's University. Within that vibrant intellectual forum, the author had the invaluable privilege of learning the deep, structural properties of integrable systems and Poisson geometry \cite{OB96, OB98, OB2007}. The profound lectures and rigorous discussions surrounding the algebraic intricacies of the Nijenhuis tensor \cite{AN1951}, the surface-forming geometry of the Haantjes tensor \cite{H1955}, and the underlying calculus of the Schouten bracket \cite{JS1940} and Schouten tensor \cite{OB2007, JS1954} (also known as the Magri–Morosi concomitant \cite{VMP2025, MM1984}) left an indelible mark on his academic path. 

He also extends his thanks to Rui Fernandes for illuminating discussions at the conference ``Symmetry, Invariants, and their Applications – A Celebration of Peter Olver's 70th Birthday.'' These conversations regarding recent advances in Poisson geometry, as well as the sharing of reference \cite{CFM2021}, were invaluable to the completion of this project. 

The second author also warmly thanks his collaborator Richard Vale, whose insightful discussions introduced a new perspective: treating the three studied growth regimes as solutions to certain optimization problems \cite{SV2026}. 

Additionally, he acknowledges with gratitude useful discussions with Davide La Torre and Simone Marsiglio on the properties and applications of the $S$-shaped production function analyzed in this paper. 

He is also grateful to Raffaele Vitolo for pointing out an important chronological detail regarding the history of the bi-Hamiltonian formalism, which  helped improve the historical accuracy and presentation of this work.

Last but not least, he wishes to thank his former graduate students Marzieh Palizdar, Timothy Power, and Kunpeng Wang for their  intellectual curiosity,  enthusiasm, and hard work, which helped launch this project.

This research was supported in part by an AIO Research Grant from the Faculty of Science at Dalhousie University.


\pdfbookmark[1]{References}{ref}
\LastPageEnding

\end{document}